\documentclass[preprint, 3p, authoryear ]{elsarticle}

\usepackage{amssymb}
\usepackage{amsmath}
\usepackage[colorlinks, linkcolor=blue, citecolor=blue, urlcolor=blue]{hyperref}
\usepackage{booktabs} 
\usepackage{hhline} 
\usepackage{booktabs} 
\usepackage{array}
\usepackage{graphicx}
\usepackage{booktabs}
\usepackage{multirow}

\journal{International Journal of Forecasting}

\begin{document}

\begin{frontmatter}


\title{A Hybrid Strategy for Probabilistic Forecasting and Trading of Aggregated Wind-Solar Power: Design and Analysis in HEFTCom2024}


\author[1]{Chuanqing Pu}
\ead{sashabanks@sjtu.edu.cn}
\author[1]{Feilong Fan\corref{cor1}}
\ead{feilongfan@sjtu.edu.cn}
\author[1]{Nengling Tai}
\ead{nltai@sjtu.edu.cn}
\author[1]{Songyuan Liu}
\ead{arronal@sjtu.edu.cn}
\author[1]{Jinming Yu}
\ead{yujinming@sjtu.edu.cn}

\affiliation[1]{organization={College of Smart Energy, Shanghai Jiao Tong University},
            city={Shanghai},
            postcode={201100}, 
            country={China}} 

\cortext[cor1]{Corresponding author}

\begin{abstract}
    Obtaining accurate probabilistic energy forecasts and making effective decisions amid diverse uncertainties are routine challenges in future energy systems. This paper presents the winning solution of team GEB, which ranked 3rd in trading, 4th in forecasting, and 1st among student teams in the IEEE Hybrid Energy Forecasting and Trading Competition 2024 (HEFTCom2024). The solution provides accurate probabilistic forecasts for a wind-solar hybrid system, and achieves substantial trading revenue in the day-ahead electricity market. Key components include: (1) a stacking-based approach combining sister forecasts from various Numerical Weather Predictions (NWPs) to provide wind power forecasts, (2) an online solar post-processing model to address the distribution shift in the online test set caused by increased solar capacity, (3) a probabilistic aggregation method for accurate quantile forecasts of hybrid generation, and (4) a stochastic trading strategy to maximize expected trading revenue considering uncertainties in electricity prices. This paper also explores the potential of end-to-end learning to further enhance the trading revenue by shifting the distribution of forecast errors. Detailed case studies are provided to validate the effectiveness of these proposed methods. Code for all mentioned methods is available for reproduction and further research in both industry and academia.

\end{abstract}
\end{frontmatter}

\section{Introduction}
\label{sec1}

Accurately forecasting renewable energy production and enabling timely decision-making are crucial for the transition to resilient, low-carbon power systems \citep{Background1,Background2,Background3}. To bridge the gap between academic research and industrial practice, the Hybrid Energy Forecasting and Trading Competition 2024 (HEFTCom2024) challenges participants to provide quantile forecasts for a hybrid power plant which consists of the Hornsea 1 offshore wind farm and solar capacity in East England, and to trade the energy in a day-ahead electricity market. 

The competition is divided into two tracks: forecasting and trading. This paper details the methodology of team GEB in HEFTCom2024 in both tracks. In the final leaderboard, we ranked as the best-placed student team in both tracks, achieving 3rd place in the trading track and 4th place in the forecasting track, respectively. Notably, we are also the only top-3 team in the trading track whose methodology is built without the use of external data that beyond the competition scope. 

Our performance can be attributed to two key factors. On one hand, the primary wind and solar power forecasts are provided by Gradient Boosting Decision Trees (GBDTs), which is a popular ensemble learning approach that has demonstrated strong performance in previous energy forecasting competitions, such as Global Energy Forecasting Competitions (GEFCom) 2017 and GEFCom2014 \citep{GEFcom1, GEFcom2, GEFcom3,GEFcom4,GEFcom5}. We also incorporated several practical techniques including feature engineering with spatial and temporal meteorological information, hyperparameter optimization and ensemble learning. These techniques are widely recognized as effective strategies in forecasting practice \citep{feature-engi, hyper-opt}. On the other hand, building on these proven techniques, we designed series tailored approaches to address the specific challenges posed by HEFTCom2024. To clarify the motivation behind our approaches, the following introduce these challenges or characteristics of HEFTCom2024, along with our corresponding solutions.

\textbf{(1) Multiple NWP sources.} To reduce the forecast variance, we developed a stacking-based ensemble method to combine sister forecasts trained on two different NWP sources provided by HEFTCom2024. Stacking is a widely used ensemble learning technique that improves forecasting performance by leveraging the complementary strengths of diverse base models \citep{stacking_review}. Stacking typically involves integrating models with different structures, such as combining Support Vector Regression (SVR) and Random Forest for day-ahead wind power forecasting \citep{stacking1}, or combining Long Short 
Term Memory (LSTM) and Informer networks for solar power forecasting \citep{stacking2}. In contrast to heterogeneous ensembles, combining sister forecasts is also demonstrated effective in power load forecasting tasks \citep{sister-forecast-load}, where sister forecasts refer to predictions generated from a series of models that share similar architectures but are trained using different input sources or variable selections \citep{sister-forecast}. In our solution, we stack sister forecasts trained on different NWP sources to exploit their spatial complementarity, as we observed discrepancies in grid point density and coverage area in the NWP data provided for the Hornsea 1 region, as shown in Section \ref{sec2}. Rather than directly merging multiple NWP sources into a single forecasting model, stacking sister models offers an additional advantage—it mitigates the risk of missing data from one NWP source during online testing. Since the two NWPs cover different spatial regions, embedding both into a single model would make the model vulnerable to data gaps if one NWP source becomes unavailable. In contrast, with a stacked ensemble, one sister model can still generate forecasts even when the other fails.

\textbf{(2) Solar distribution shift.} The solar capacity in East England was observed from 2,609 MWp to 2,741 MWp as of February 19, 2024. Since the offline dataset only reflects capacities up to 2,609 MWp, the forecasting models solely trained on offline dataset become less effective at online test-set. Distribution shift is a typical challenge in forecasting practice. Common methods to address distribution shifts include oversampling, transfer learning, and post-processing. Oversampling replicates limited new samples to construct an augmented training set, which can alleviate overfitting to outdated data but may disrupt temporal continuity and lead to overfitting on the limited new samples or being overwhelmed by historical samples \citep{oversample1, oversample2}. Transfer learning aim to align the model outputs with the new data distribution through fine-tuning \citep{transfer_learning1} or domain adaptation \citep{transfer_learning2}, but the structure of decision trees does not support typical transfer learning paradigms. Post-processing operates directly on model outputs, which is effective for linear shifts and does not require retraining the model. While they assume the existence of a known and simple shift pattern, such assumptions are valid in our case: the increase in solar capacity is both known and induces a near-linear expansion in power generation. We combined online learning with post-processing to dynamically fit the new distribution. Given the limited online samples, we adopted a polynomial post-processing model trained using LASSO-penalized quantile regression \citep{LASSO} for calibration, which prevents overfitting of higher-order terms by penalizing the model complexity. While wind power forecasts may also be subject to distributional changes, such shifts tend to be less systematic and harder to characterize without prior knowledge. Applying nonlinear corrections to wind power forecasts under sparse online data conditions will introduce additional risks of overfitting. Consequently, we restrict the use of online post-processing to solar power forecasts in the final solution.

\textbf{(3) Requirement to provide quantile forecasts of total generation.} Directly predicting the distribution of total generation would hinder targeted post-processing, as Hornsea 1 encountered planned outage that were not present in the offline dataset, and the distribution of solar power generation also exhibited a shift. Therefore, forecasting wind and solar power separately is necessary and also help reduces the capacity of the forecasting model. While summing the forecast results quantile by quantile offers simplicity, it theoretically sacrifices the accuracy. Discrete convolution can be used to combine multiple cumulative distribution functions (CDFs) to obtain the distribution of the sum \citep{aggregate1}. Copula-based discrete convolution methods are further researched to aggregate the uncertainties of wind power \citep{aggregate2} and load demand \citep{aggregate3} in power system. Inspired by these theoretical frameworks and applications, we developed an aggregated probabilistic forecasting approach by first predicting the CDF of wind and solar power separately and then reconstructing the distribution of total generation via discrete convolution. The CDF is estimated by learning the dense quantiles. This approach allows for an accurate estimation of the overall generation distribution in a parameter-free manner. Furthermore, this approach benefits from powerful models like LightGBM \citep{LightGBM} or CatBoost \citep{CatBoost}, which directly support learning the pinball loss.

\textbf{(4) Requirement to trade energy in the day-ahead electricity market}. The trading revenue is influenced by the uncertainties in day-ahead prices, imbalance prices, and power generation of the hybrid power plant. Due to limited information about the market and unknown market imbalances, price forecasting is challenging. However, statistically, the mean of the price spread follows a seasonal pattern throughout the day. To utilize this regularity, we developed a stochastic trading method based on uncertain price spread and deterministic power forecasts to maximize the expected trading revenue. We derived a revenue loss from the inaccuracies in both power forecasts and price spread forecasts to guide the training of the deterministic power forecasting model. This approach provides a simple and robust strategy that avoids the need for training price forecasting models. Additionally, we propose a further enhancement to improve the trading revenue through end-to-end learning (E2EL). E2EL refers to designing forecasting models with a focus on downstream decision-making tasks rather than solely minimizing statistical errors. E2EL is explored in related literature for unit commitment \citep{E2E_UC}, economic dispatch \citep{E2E_ED}, and energy arbitrage \citep{E2E_ESS} tasks. The key to achieving superior decision-making through E2EL lies in leveraging the asymmetric impact of forecast errors on downstream benefits, which is overlooked in accuracy-oriented forecasting. We also discovered this characteristic of error asymmetry in the market rules set by HEFTCom2024, and case studies demonstrate that E2EL can enhance trading revenue by adjusting the distribution preferences of forecast errors.

The remainder of this paper is organized as follows. Section \ref{sec2} formulate the tasks of HEFTCom2024. Section \ref{sec3} and Section \ref{sec4} details the methodology in the forecasting and trading track. Section \ref{sec5} presents case studies to validate the effectiveness of the methodology. Section \ref{sec6} discusses the trick and lessons learned from HEFTCom2024 and explores the potential to further enhance the revenue. Section \ref{sec7} concludes the paper.

\section{Problem Statement}
\label{sec2}

\subsection{Tasks in Forecasting Track}
The forecasting track requires participants to provide quantile forecasts from 10\% to 90\% in 10\% intervals for the total generation. The total generation is the sum of wind and solar power, expressed as
\begin{equation}
    y_t = y_{\text{w}.t} + y_{\text{s}.t},
\end{equation}
where $y_{\text{w}.t}$ and $y_{\text{s}.t}$ denote the actual wind and solar power generation, respectively.

The forecasts are submitted daily at a half-hour resolution, resulting in a total of $48 \times 9$ predictions. Let $\hat{y}_{\tau,t}$ denote submission results from the participants, i.e., the predicted total generation at quantile $\tau$ and time $t$, and $y_t$ denote the actual total generation at time $t$. The forecasts will be evaluated using the pinball loss
\begin{equation}
    L_{\tau}(y_t, \hat{y}_{\tau,t}) = (\hat{y}_{\tau,t} - y_t) \left[ \tau \cdot \mathbb{I}(\hat{y}_{\tau,t} < y_t) - (1 - \tau) \cdot \mathbb{I}(\hat{y}_{\tau,t} \geq y_t) \right],
\end{equation}
where $\mathbb{I}(\cdot)$ is the indicator function.

The final forecasting score is calculated as the average pinball loss across all elements in the quantiles set $\mathcal{Q}= \{0.1, 0.2, \ldots, 0.9\}$ and all half-hour periods, which is defined as
\begin{equation}
    \label{eq:MPL}
    S_{\text{forecasting}}=\frac{1}{|\mathcal{Q}|} \sum_{\tau \in \mathcal{Q} } [\frac{1}{48}\sum_{t=1}^{48} L_{\tau}(y_t, \hat{y}_{\tau,t})].
\end{equation}

\subsection{Tasks in Trading Track}
The trading track is based on Great Britain's wholesale electricity market. Bidding decisions are made only in the day-ahead market. The participants are required to submit the energy bidding results $\hat{e}_t$ for each half-hour period $t$. The goal of the trading track is to maximize the following score
\begin{equation}
    \label{revenue}
    S_{\text{trading}}=\sum_{t=1}^{48} \hat{e}_t\cdot\pi_{\text{DA},t}+(y_t-\hat{e}_t)\cdot\pi_{\text{SS},t}-0.07(y_t-\hat{e}_t)^2,
\end{equation}
where $\pi_{\text{DA},t}$, $\pi_{\text{SS},t}$ denote the day-ahead and imbalance prices, respectively. 

\subsection{Provided Data}
\begin{figure}[h]
    \centering
    \includegraphics[width=0.4\textwidth]{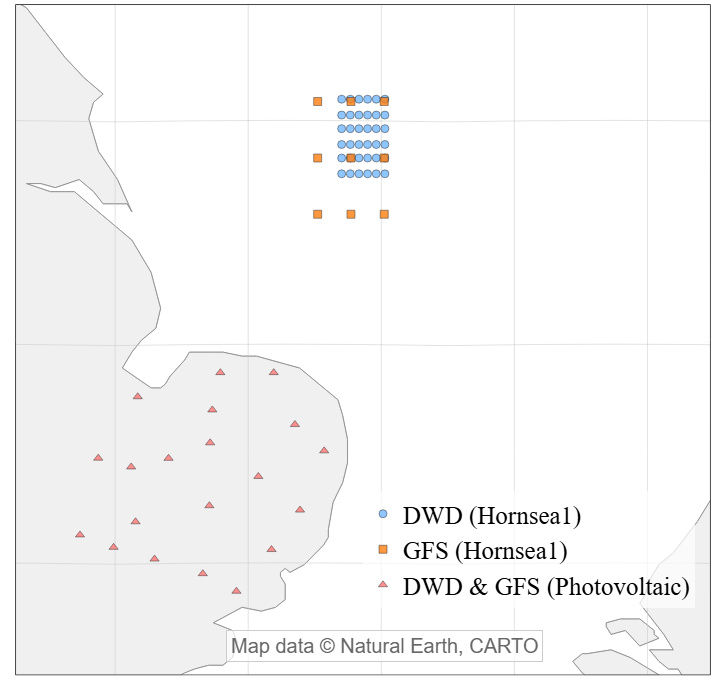}
    \caption{Grid coordinates of historical meteorological data for Hornsea 1 and East England PV plants.}
    \label{metrodata}
\end{figure}

HEFTCom2024 provides two meteorological data sources for forecasting: Deutscher Wetterdienst (DWD) and the Global Forecast System (GFS). The grid coordinates of the historical meteorological data for Hornsea 1 and East England photovoltaic (PV) plants are shown in Figure \ref{metrodata}. For Hornsea 1, 36 coordinates are provided from the DWD, and 9 coordinates are provided from the GFS. For the solar power plants, 20 coordinates are provided from both DWD and GFS sources. The data types include wind speed and wind direction at 100m and 10m heights, temperature, solar radiation, cloud cover, and other meteorological data, with hourly resolution. The data are corrected every six hours, resulting in multiple corrected data points for the same valid time. The historical power data for Hornsea 1 and East England PV plants are given at half-hour intervals \citep{data}. During the online testing phase, the wind power data for Hornsea 1 is typically updated with a one-week lag, while the solar power plant data is updated with a one-day lag. HEFTCom2024 also provides day-ahead and imbalance price data in Great Britain's wholesale electricity market for the trading track. The aforementioned data span from September 20, 2020, to January 18, 2024. The evaluation period for the competition is from February 20, 2024, to May 19, 2024.

\section{Methodology in Forecasting Track}
\label{sec3}
\subsection{Overview}
\begin{figure*}[h]
    \centering
    \includegraphics[width=1\textwidth]{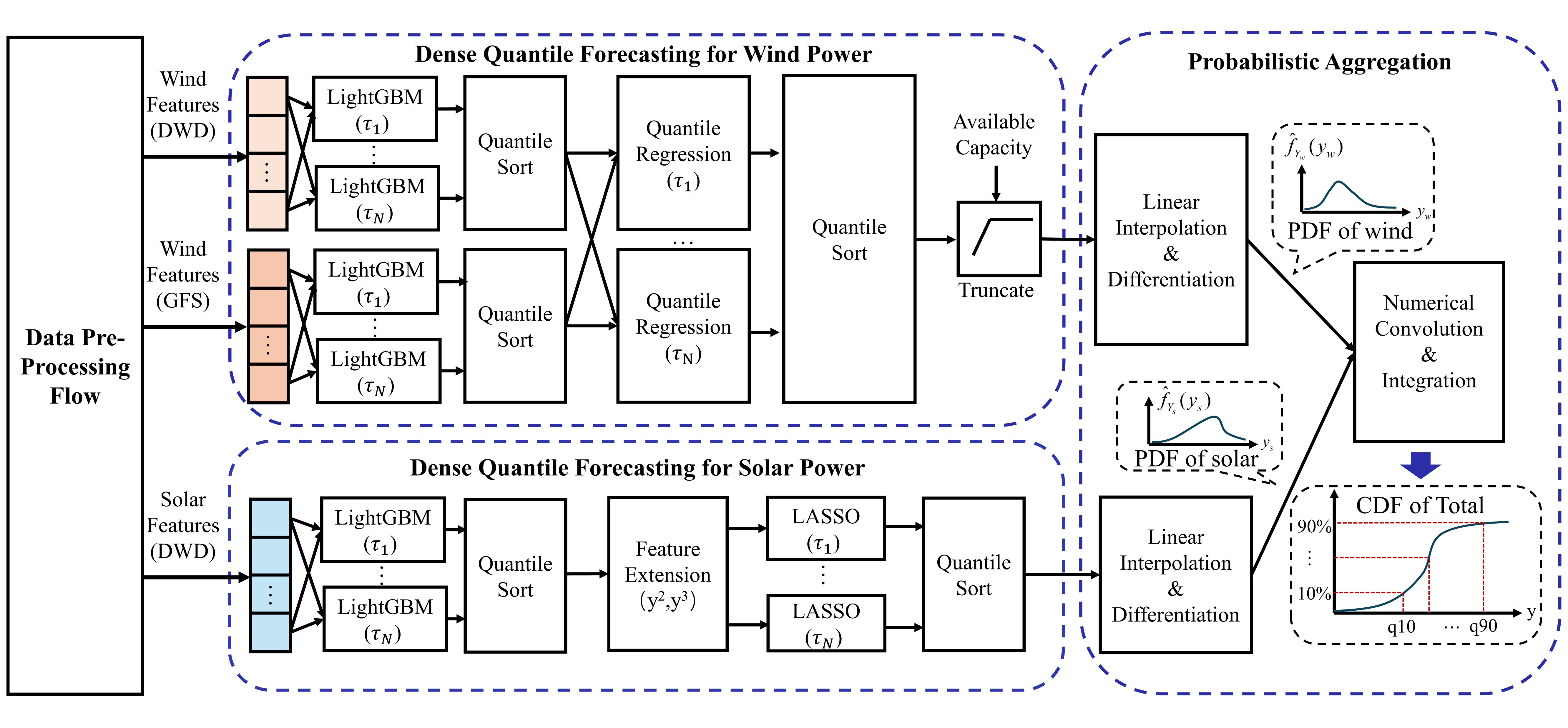}
    \caption{Overview of the methodology in the forecasting track. It consists of four modules: data pre-processing, wind power forecasting, solar power forecasting, and probabilistic aggregation. The final forecast result is the CDF of total generation, as shown in the curve plot in the lower right corner. The quantile forecasts from 10\% to 90\% in 10\% increments are obtained from the inverse function of the CDF.
    }
    \label{overview}
\end{figure*}

The overall framework of the developed methodology is illustrated in Figure \ref{overview}. The methodology consists of four main modules: (1) Data Pre-processing Pipeline, (2) Dense Quantile Forecasting for Wind Power, (3) Dense Quantile Forecasting for Solar Power, and (4) Probabilistic Aggregation. The core idea is to model the probability density function (PDF) of wind and solar power through dense quantile forecasts, which are then aggregated to fit the distribution of total generation to obtain the target quantiles. LightGBM is employed as the primary model in the dense quantile forecasting for wind and solar power modules, as it is a popular tree-based model notably demonstrated in the M5 forecasting competition \citep{M5}. LightGBM is also recognized for its high training efficiency. According to our observations in offline experiments, the training time for a single quantile forecasting model is less than 30 seconds on the full dataset provided by HEFTCom2024 (i914900k CPU, 96GB memory). This choice helps improve the efficiency of pre-competition development and debugging processes, and is particularly effective in training multiple quantile regression models for fitting the CDF of wind and solar power. The details of each module are elaborated in Sections \ref{datapreprocess}, \ref{windpower}, \ref{solarpower}, and \ref{totalpower}.

\subsection{Data Pre-processing Pipeline}
\label{datapreprocess}
\begin{figure*}[!ht]
    \centering
    \includegraphics[width=1\textwidth]{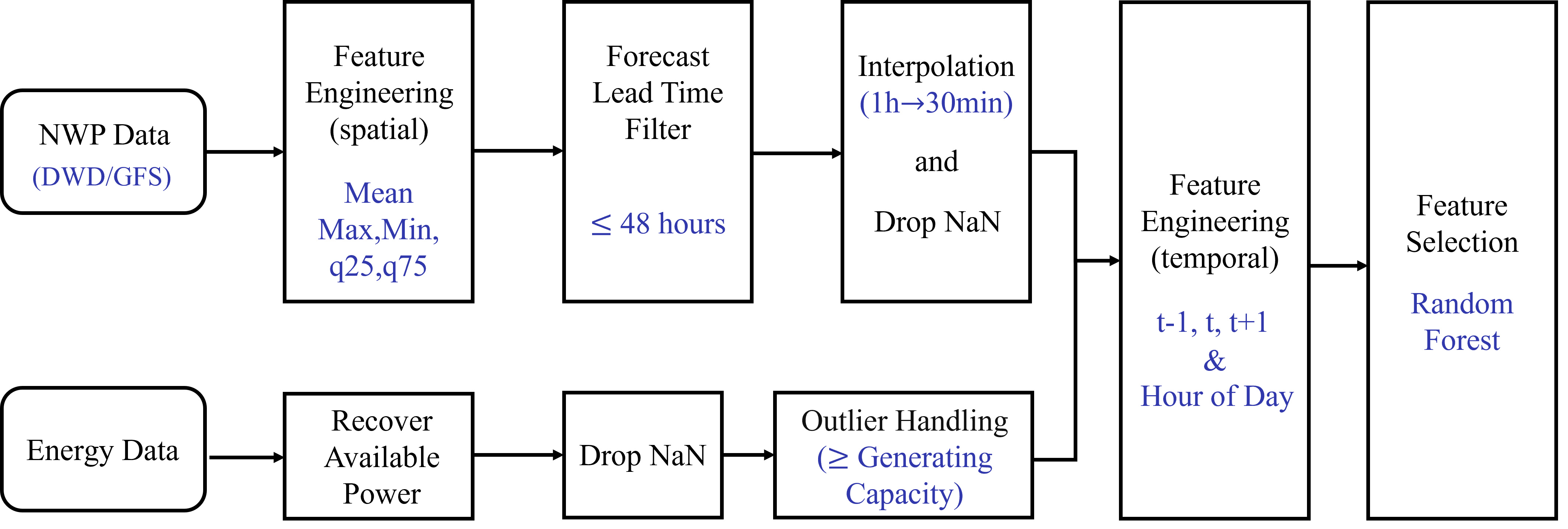}
    \caption{Data pre-processing pipeline for the HEFTCom2024 data.}
    \label{fig1}
\end{figure*}

The data pre-processing pipeline is illustrated in Figure \ref{fig1}. We begin by processing the NWP data from the DWD and GFS models through spatial feature extraction, which reduces the dimensionality of the dense grid data. The extracted features include the mean, maximum, minimum, 25th percentile, and 75th percentile of the meteorological variables. Subsequently, we retain only NWP data with forecast lead time (FLT) within 48 hours, as the latest available FLT at the time of daily forecast submissions typically range from 23 to 47 hours. Including NWP data with FLT less than 23 hours can enhance forecast accuracy, while excluding those beyond 48 hours helps mitigate noise. Linear interpolation is employed to resample the NWP data to a half-hourly resolution. For the energy data, we reconstruct the actual available wind power based on the Bid Offer Acceptances (BOA) from the system operator. Subsequently, we perform outlier treatment on the power generation data from Hornsea 1 and solar power plants, which includes removing NaN values and filtering power values that exceed generation capacity. Once the energy and meteorological data are pre-processed, we align them to create the modeling dataset. In temporal feature engineering, we select time-shifted features for both past and future half-hour intervals. To determine the most relevant features, Random Forest (RF) \citep{RF} is used to rank feature importance. For wind power forecasting, 9 features are selected by combining the maximum, mean, and minimum wind speeds at 100 meters in the target region with lagged, leading, and current time steps. Similarly, for solar power forecasting, 13 features are selected by combining the maximum, mean, and minimum values of solar downward radiation and cloud cover in the target region, along with lagged, leading, and current time steps, as well as the hour-of-day. It is noteworthy that the final feature set is not strictly determined by the RF rankings. Manual selection is also performed based on train-test performance to determine the final features.

\subsection{Wind Power Forecasting based on Stacked Sister Models}
\label{windpower}
Given the discrepancies in grid point density and spatial coverage between the NWP from GFS and DWD, which are shown in Figure \ref{metrodata}, leveraging the complementarity of the two NWP sources helps enhance forecasting performance. Specifically, two different series of models using DWD and GFS NWP data are trained, respectively. In each series, to fit the CDF, a separate forecasting model will be trained for each quantile $\tau$, where $\tau \in \mathcal{B}=\{0.1\}\cup\{1,2,...,99\}\cup\{99.9\}$. The final wind power forecast is obtained by stacking the outputs of the two series of models. The detailed methodology is described as follows.

Take DWD as an example, given the dataset $(\mathbf{x}_{\text{w}.t},y_{\text{w}.t}) \in \mathcal{W}_{\text{DWD}}$, where $\mathbf{x}_{\text{w}.t} \in \mathbb{R}^{9}$ denotes the features and $y_{\text{w}.t} \in \mathbb{R}$ denotes the actual wind power, for quantile $\tau$, the LightGBM model $F_{\text{DWD}.\tau}$ is represented as

\begin{equation}
    \begin{aligned}
    F_{\text{DWD}.\tau}(\mathbf{x}_{\text{w}.t}) = \sum_{m=1}^{M} \eta \cdot h_m(\mathbf{x}_{\text{w}.t}),
    \end{aligned}
\end{equation}
where $h_m(\mathbf{x}_{\text{w}.t})$ denotes the prediction of the $m$-th tree in the model, and $\eta$ denotes the learning rate. The gradient boosting algorithm fits the negative gradient of the previous model by iteratively constructing new trees. Let $\hat{F}_{\text{DWD}.\tau}^{(m)}(\mathbf{x}_{\text{w}.t})$ denote the model in the $m$-th iteration, the negative gradient of the current model is
\begin{equation}
    \begin{aligned}
        r_{\text{DWD}.\tau,m}=-\frac{\partial L_{\tau}(y_{\text{w}.t},\hat{F}_{\text{DWD}.\tau}^{(m-1)}(\mathbf{x}_{\text{w}.t}))}{\partial \hat{F}_{\text{DWD}.\tau}^{(m-1)}(\mathbf{x}_{\text{w}.t})} \quad t=1,2,...,|\mathcal{W}_{\text{DWD}}|,
    \end{aligned}
\end{equation}
where the new tree $h_{m}$ is trained to minimize the loss function
\begin{equation}
    \begin{aligned}
    h_{m}(\mathbf{x}_{\text{w}.t})= \underset{h}{\text{argmin}} \sum_{t=1}^{|\mathcal{W}_{\text{DWD}}|} L_{\tau}(y_{\text{w}.t}, \hat{F}_{\text{DWD}.\tau}^{(m-1)}(\mathbf{x}_{\text{w}.t})+h(\mathbf{x}_{\text{w}.t})).
    \end{aligned}
\end{equation}

The updated model is obtained by adding the new tree to the current model, expressed as
\begin{equation}
    \begin{aligned}
    \hat{F}_{\text{DWD}.\tau}^{(m)}(\mathbf{x}_{\text{w}.t})=\hat{F}_{\text{DWD}.\tau}^{(m-1)}(\mathbf{x}_{\text{w}.t})+\eta \cdot h_{m}(\mathbf{x}_{\text{w}.t}).
    \end{aligned}
\end{equation}

Building on this, LightGBM introduces improvements such as the histogram algorithm, leaf-wise best split algorithm, and parallel learning, which are not detailed here but can be found in the original paper \citep{LightGBM}. The final wind power forecast is the stacked output of the two models, which can be represented as
\begin{equation}
    \begin{aligned}
    \hat{y}_{\text{w}.\tau,t}= g_{\omega_{\tau}}(F_{\text{DWD}.\tau}(\mathbf{x}_{\text{w}.t}), F_{\text{GFS}.\tau}(\mathbf{x}_{\text{w}.t}^{'})),
    \end{aligned}
\end{equation}
where $(\mathbf{x}_{\text{w}.t}^{'},y_{\text{w}.t}) \in \mathcal{W}_{\text{GFS}}$, and $g_{\omega_{\tau}}$ is the quantile regression model trained to combine the outputs of the two base models. The optimal weights $\omega_{\tau}^*$ are determined by minimizing the pinball loss on the aligned training set, which is defined as
\begin{equation}
    \begin{aligned}
    \omega_{\tau}^* = \underset{\omega_{\tau}}{\text{argmin}}\sum_{t=1}^{|\mathcal{W}_{\text{DWD}}|} L_{\tau}(y_{\text{w}.t}, \hat{y}_{\text{w}.\tau,t}).
    \end{aligned}
\end{equation}

A special event that affects wind power forecasting is the planned outage of Hornsea 1 during the competition, which will limit the available capacity of wind power. This event has not been reflected in the offline dataset before. We design a quantile-wise truncation model to correct the wind power forecast results based on the available capacity of Hornsea 1, which is reported in the REMIT of ELEXON BSC website \citep{REMIT}. The corrected wind power forecast $\hat{y}_{\text{w}.\tau,t}^{'}$ is obtained by
\begin{equation}
    \hat{y}_{\text{w}.\tau,t}^{'} =\min(\hat{y}_{\text{w}.\tau,t}, c_{\tau} \cdot Q_{\text{Hornsea-1}}),
\end{equation}
where $c_{\tau}$ denotes the truncation coefficient, and $Q_{\text{Hornsea-1}}$ denotes the available capacity of Hornsea 1. Since the observed actual maximum wind power is usually lower than the available capacity reported on ELEXON BSC website, a truncation coefficient is set. The truncation coefficient is fine-tuned based on the observed online dataset, which can be represented as
\begin{equation}
    \begin{aligned}
        \{c_{\tau}^*\} = \underset{\{c_{\tau} \in [0,1]\}}{\text{argmin}} \sum_{t=1}^{|\mathcal{W}_{\text{latest}}|} L_{\tau}(y_{\text{w}.t}, \hat{y}_{\text{w}.\tau,t}^{'}), \\
        \text{subject to } c_{\tau_i} \geq c_{\tau_j}, \quad \forall \tau_i > \tau_j
    \end{aligned}
\end{equation}
where $\mathcal{W}_{\text{latest}}$ denotes the latest online dataset. In the final solution, the truncation coefficient is dynamically adjusted between 0.9 and 1.0. It is also necessary to ensure that the truncation coefficient increases with higher quantiles $\tau$ to avoid quantile misalignment.

\subsection{Solar Power Forecasting with Online Post-processing}
\label{solarpower}
\begin{figure}[!ht]
\centering
\hspace{-0.5cm}
\includegraphics[width=0.45\textwidth]{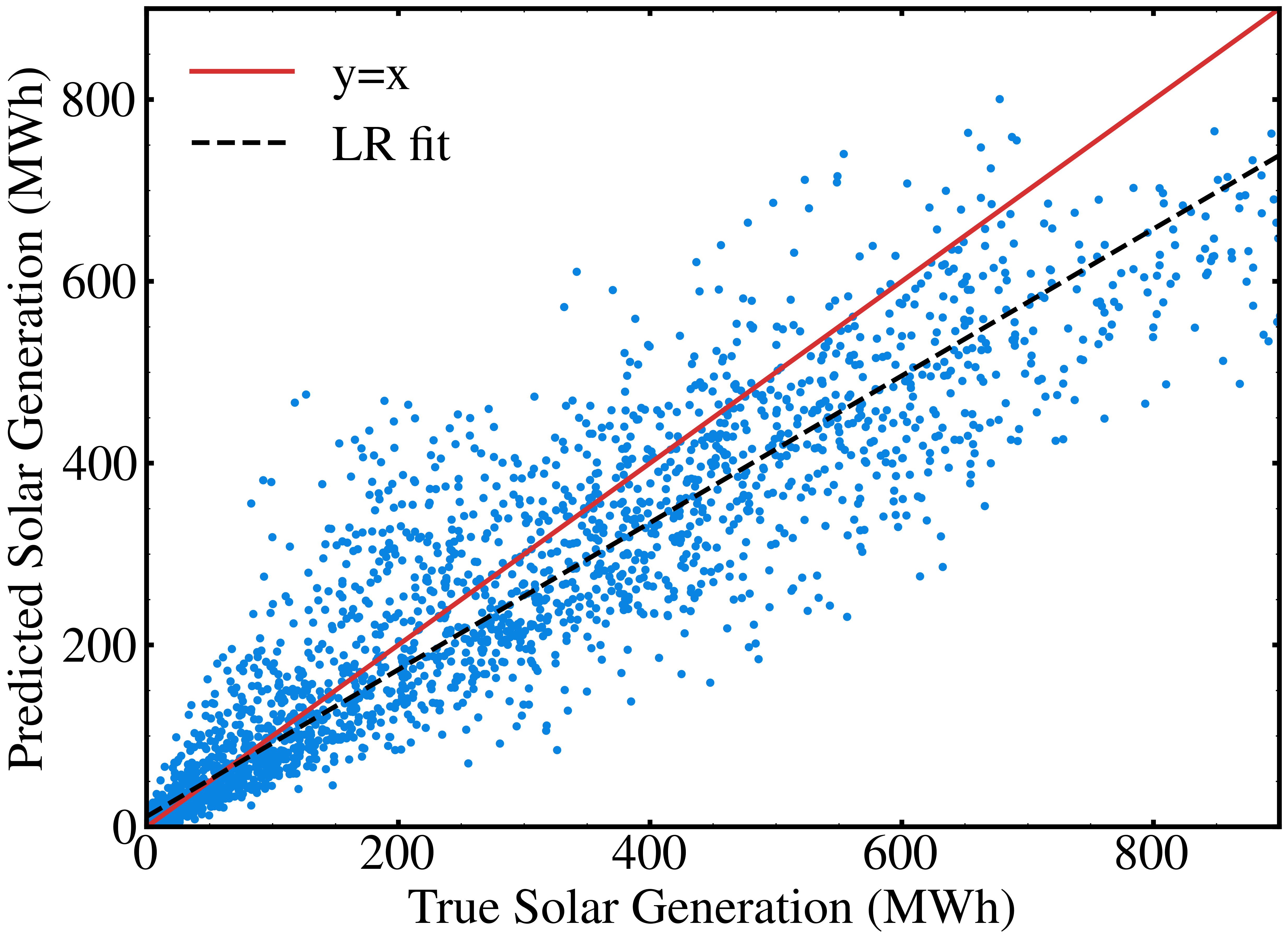}
\caption{The test results of solar power forecasting models trained on historical data. The line “y=x” denotes the point set that the predicted value equals the actual value. The line “LR fit” denotes the linear regression fit of the point set.}
\label{fig2}
\end{figure}

The primary model for solar power forecasting is similar to that for wind power forecasting, with LightGBM used to predict quantiles $\tau$. Due to the increased solar capacity, which can be determined through the API provided by the competition, the solar power forecasting models trained solely on historical data are not fully applicable. Figure \ref{fig2} illustrates the test results of the original forecasting model on the online dataset of HEFTCom2024. The results indicate that models trained only on historical offline dataset tend to underestimate the solar power and fail to capture instances of high solar power. 

Since the limited performance of the offline model is not due to data randomness but rather caused by distribution shift, we propose an online post-processing model to correct such distribution shifts based on limited online samples. For each quantile $\tau$ in $t$-th sample, let $\mathcal{S}$ denote the latest online dataset of solar power, and $\hat{y}_{\text{s}.\tau,t}$ denote the original solar power forecast from the LightGBM models, the corrected solar power forecast $\hat{y}_{\text{s}.\tau,t}^{'}$ is obtained by a polynomial post-processing model
\begin{equation}
    \hat{y}_{\text{s}.\tau,t}^{'}=\beta_1 \cdot \hat{y}_{\text{s}.\tau,t}+\beta_2\cdot (\hat{y}_{\text{s}.\tau,t})^2+ \beta_3\cdot (\hat{y}_{\text{s}.\tau,t})^3 \quad t=1,...,|\mathcal{S}|, \tau \in \mathcal{B},
\end{equation}
where the parameters $\beta_1$, $\beta_2$, and $\beta_3$ denote the coefficients of the polynomial model. 

Due to the limited number of samples at the current stage, the model complexity must be controlled. We employ a LASSO-penalized quantile regression method to train the post-processing model. For each quantile $\tau$, the objective of the LASSO-penalized regression is to minimize the following loss function
\begin{equation}
    \underset{\beta_1,\beta_2,\beta_3}{\text{min}}
     \quad \sum_{t=0}^{|\mathcal{S}|} L_{\tau}(y_{\text{s}.t}, \hat{y}_{\text{s}.\tau,t}^{'})+\lambda\sum_{j=1}^{3}|\beta_j|,
\end{equation}
where $y_{\text{s}.t} \in \mathbb{R}$ denotes the actual solar power, and $\lambda$ is the L1-regularization parameter.

The latest solar power samples are collected daily, and the regression model is retrained to update $\beta_1$, $\beta_2$, and $\beta_3$. To tune the penalty $\lambda$, the solar generation and corresponding 00:00 UTC NWP in the online dataset are split 60/40 into training and validation sets. The LASSO-penalized quantile regression is fit on the training data, and $\lambda$ is chosen to minimize the quantile loss on the validation set. Once the optimal $\lambda$ is identified, the model is retrained on the full dataset to obtain the final coefficients.

\subsection{Aggregated Quantile Forecasting}
\label{totalpower}
To obtain quantile forecasts for total power generation, it is necessary to derive the CDF of total generation conditioned on covariates $\mathbf{x}$, where $\mathbf{x}$ represents the combined features used for wind and solar power, i.e., $\mathbf{x} = [\mathbf{x}_{\text{w}}^T, \mathbf{x}_{\text{s}}^T]^T$. The primary approach involves reconstructing the PDF and CDF for wind and solar power based on their respective quantile forecasts. These reconstructed distributions are then numerically convolved to derive the PDF and CDF of total generation. A key assumption for the numerical convolution of the PDFs for wind and solar power is that they are conditionally independent given the covariates $\mathbf{x}$. This assumption has been validated through regression-based testing \citep{regression-based test}, with the results provided in the Appendix A. The proposed method is non-parametric and requires the following steps to be implemented.

Taking wind power forecasting as an example, we ignore the temporal dimension $t$ and focus on a single sample, $|\mathcal{B}|$ quantile forecasts $\hat{y}_{\text{w}.\tau_1},\dots, \hat{y}_{\text{w}.\tau_i} ,\dots, \hat{y}_{\text{w}.\tau_{|\mathcal{B}|}} (\tau_i \in \mathcal{B})$ are used to obtain the CDF $F_{y_{\text{w}} \mid \mathbf{x}_{\text{w}}}(y \mid \mathbf{x}_{\text{w}})$ of wind power through linear interpolation, expressed as
\begin{equation}
    \begin{aligned}
        F_{y_{\text{w}} \mid \mathbf{x}_{\text{w}}}(y \mid \mathbf{x}_{\text{w}}) = \sum_{i=1}^{|\mathcal{B}|} \left( \tau_i + \frac{y - \hat{y}_{\text{w}.\tau_i}}{\hat{y}_{\text{w}.\tau_{i+1}}- \hat{y}_{\text{w}.\tau_i}} (\tau_{i+1} - \tau_i) \right) \mathbb{I}(\hat{y}_{\text{w}.\tau_i} \leq y < \hat{y}_{\text{w}.\tau_{i+1}})
        + \mathbb{I}(y \geq \hat{y}_{\text{w}.\tau_N}),
\end{aligned}
\end{equation}
where $y \in \mathcal{G_{\text{w}}}= \{y_{\text{w}}^{(1)},y_{\text{w}}^{(2)},...,y_{\text{w}}^{(D_{\text{w}})}\}$ is a series of discretized points with fixed interval $\Delta y$.

The PDF of wind power is then obtained by differentiating the CDF, expressed as
\begin{equation}
    \begin{aligned}
        f_{y_{\text{w}} \mid \mathbf{x}_{\text{w}}}(y \mid \mathbf{x}_{\text{w}}) = \frac{d}{dy} F_{y_{\text{w}} \mid \mathbf{x}_{\text{w}}}(y \mid \mathbf{x}_{\text{w}}).
    \end{aligned}
\end{equation}

Similar methods are used to obtain the CDF $F_{y_{\text{s}} \mid \mathbf{x}_{\text{s}}}(y \mid \mathbf{x}_{\text{s}})$ and PDF $f_{y_{\text{s}} \mid \mathbf{x}_{\text{s}}}(y \mid \mathbf{x}_{\text{s}})$ for solar power, where $y$ also belongs to a discrete point set $\mathcal{G_{\text{s}}}= \{y_{\text{s}}^{(1)},y_{\text{s}}^{(2)},...,y_{\text{s}}^{(D_{\text{s}})}\}$ with fixed interval $\Delta y$.

Under the assumption of conditional independence between $f(y_{\text{w}} \mid \mathbf{x}_{\text{w}})$ and $f(y_{\text{s}} \mid \mathbf{x}_{\text{s}})$, the PDF of total generation can be obtained by numerically convolving the PDFs of wind and solar power, expressed as
\begin{equation}
    \begin{aligned}
        f_{y_{\text{w}}+y_{\text{s}} \mid \mathbf{x}_{\text{w}},\mathbf{x}_{\text{s}}}(y \mid \mathbf{x}_{\text{w}},\mathbf{x}_{\text{s}})=\sum_{i=1}^{D_{\text{w}}-1} f_{y_{\text{w}} \mid \mathbf{x}_{\text{w}}}(y_{\text{w}}^{(i)} \mid \mathbf{x}_{\text{w}}) \cdot f_{y_{\text{s}} \mid \mathbf{x}_{\text{s}}}(y-y_{\text{w}}^{(i)} \mid \mathbf{x}_{\text{s}}) \cdot \Delta y,
    \end{aligned}
\end{equation}
where $y$ belongs to a new domain $\mathcal{G_{\text{total}}}= \{y^{(1)},y^{(2)},...,y^{(K)}\}$ with fixed interval $\Delta y$.

The CDF of total generation is computed as the cumulative sum of its PDF, expressed as
\begin{equation}
    \begin{aligned}
        F_{y_{\text{w}}+y_{\text{s}} \mid \mathbf{x}_{\text{w}},\mathbf{x}_{\text{s}}}(y \mid \mathbf{x}_{\text{w}},\mathbf{x}_{\text{s}})=\sum_{i:y^{(i)} \leq y} f_{y_{\text{w}}+y_{\text{s}} \mid \mathbf{x}_{\text{w}},\mathbf{x}_{\text{s}}}(y^{(i)} \mid \mathbf{x}_{\text{w}},\mathbf{x}_{\text{s}}) \cdot \Delta y.
    \end{aligned}
\end{equation}

The final quantile forecast of total generation is extracted by solving the inverse CDF at the desired quantile levels
\begin{equation}
    \hat{y}_{q} = F^{-1}_{y_{\text{w}}+y_{\text{s}} \mid \mathbf{x}_{\text{w}},\mathbf{x}_{\text{s}}}(q) \quad q \in \mathcal{Q}=\{0.1,0.2,...,0.9\}.
\end{equation}

For scenarios where the distribution of solar power shows minimal variation (e.g., nighttime), the quantile forecast of total generation can be approximated as the sum of individual quantile forecasts, expressed as
\begin{equation}
    \hat{y}_{q} \approx \hat{y}_{\text{w}.q} + \hat{y}_{\text{s}.q} \quad q \in \mathcal{Q}.
\end{equation}

\subsection{Train-test Framework for Model Tuning}
We design a three-fold cross-validation framework to tune the hyperparameters of the LightGBM models. The test set is constructed to ensure consistency with the data quality encountered in online testing. Specifically, the reference time for the test set is fixed at 00:00, with a lead time ranging from 23 to 47 hours. Optuna \citep{Optuna} is used to optimize the hyperparameters of LightGBM. Optuna is an automated hyperparameter optimization framework that searches the hyperparameter space in parallel. Specifically, the type of hyperparameter and the range of values are shown in Table \ref{tab1}. The final hyperparameters are selected based on their performance on the three-fold cross-validation.

\begin{table}[!ht]
    \centering
    \caption{Hyperparameters of LightGBM that are optimized using Optuna.}
    \label{tab1}
    \begin{tabular}{ccc}
    \toprule
    \textbf{Name} & \textbf{Range of Values} & \textbf{Step} \\ 
    \midrule
    Learning Rate           & [0.01, 0.3]   &   -         \\ 
    Max Depth               & [3, 12]       &   1        \\ 
    Num Leaves              & [100, 1000]   & 100              \\ 
    Min Data In Leaf        & [200, 10000]    & 100             \\ 
    Num Estimators          & \{500,1000,2000\}   & -             \\ 
    Lambda L1               & [0, 100]            & 10             \\ 
    Lambda L2               & [0, 100]            & 10             \\ \bottomrule
    \end{tabular}
\end{table}

After obtaining the best hyperparameters, we retrain these models on the full dataset for deployment in actual testing. Since a sufficient number of quantile forecasts are required to fit the CDF, a huge number of models need to be trained. Specifically, considering forecasts of wind and solar power, two NWP data sources, and whether to train on the full dataset, a total of $N_{\text{NPWs}}\times N_{\text{full}}\times N_{\text{s}\&\text{w}}\times|\mathcal{B}|=808$ models need to be trained, where $N_{\text{NPWs}}=2$, $N_{\text{full}}=2$, $N_{\text{s}\&\text{w}}=2$, and $|\mathcal{B}|=101$. Therefore, only the hyperparameters of the models for the 50\% quantile are optimized as representatives.

\section{Methodology in Trading Track}
\label{sec4}
\subsection{Analysis of Electricity Prices}
\label{sec41}
For a given trading day, define the price spread $\pi_{\text{D},t}=\pi_{\text{DA},t}-\pi_{\text{SS},t}$, Eq. \eqref{revenue} can be rewritten as
\begin{equation}
    S_{\text{trading}} = \sum_{t=1}^{48} \pi_{\text{SS},t}y_t+\pi_{\text{D},t}\hat{e}_t-0.07(y_t-\hat{e}_t)^2.
\end{equation}
Excluding the constant term, the remaining revenue influenced by bidding decisions is solely related to the actual power generation $y_t$ and price spread $\pi_{\text{D},t}$. However, forecasting price spread $\pi_{\text{D},t}$ poses a significant challenge. For day-ahead prices, the lack of comprehensive information of the market participants and the absence of load and renewable generation data of this market make the forecasting task difficult. For real-time prices, they are mainly affected by market imbalances that are almost impossible to predict at day-ahead stage. Moreover, the latest available electricity price data in online test is typically delayed by 4 to 5 days, which makes it challenging to adopt extrapolation methods for price spread forecasting.

\begin{figure}[!h]
\centering
\hspace{-0cm}
\includegraphics[width=0.55\textwidth]{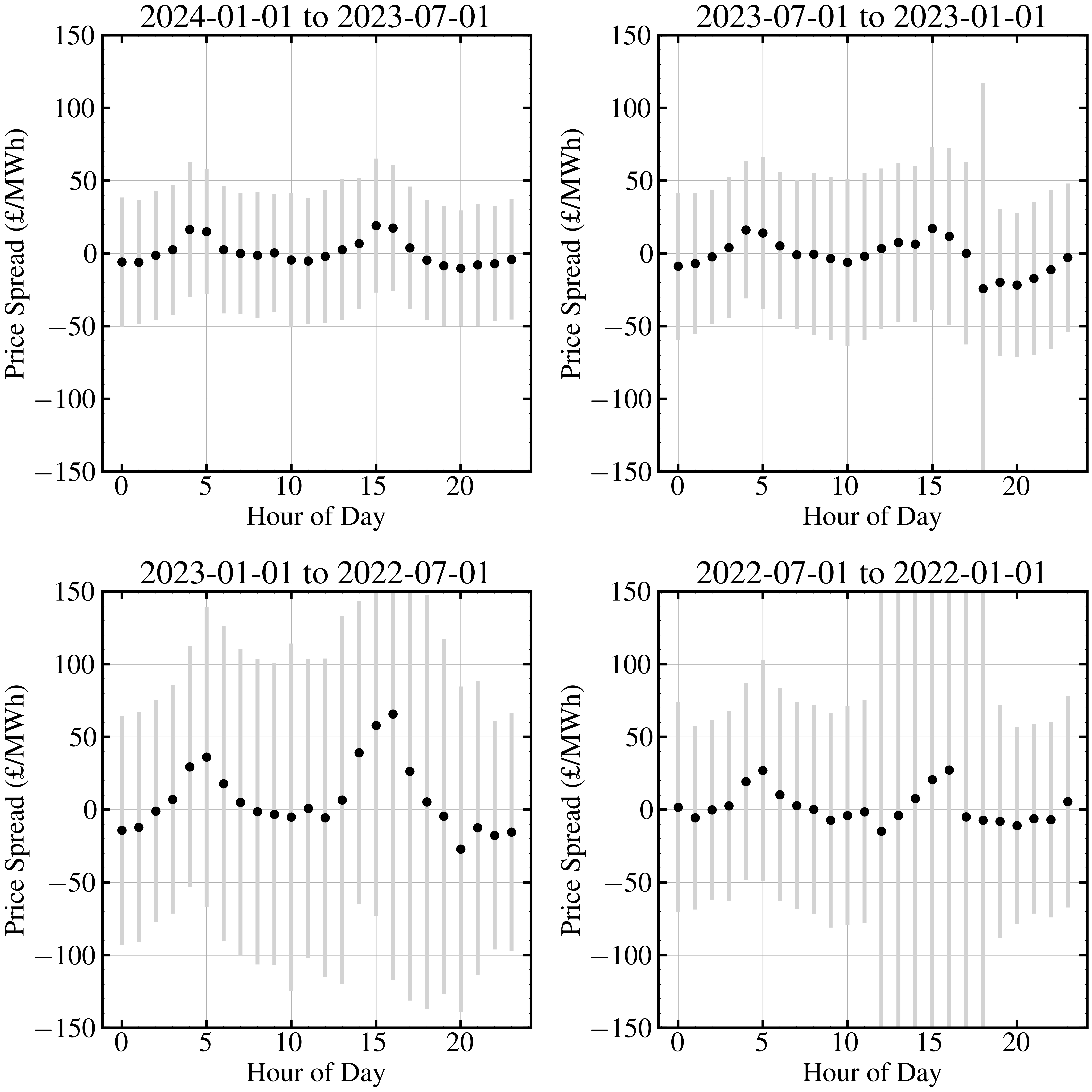}
\caption{The mean and variability of the price spread at different hour-of-day in various periods. Each subplot represents the statistical data of a half-year period. The x-axis represents different hours of the day. The markers represent the mean of the price spread at each hour, while the bars represent the standard deviation of the price spread at each hour.
}
\label{fig5}
\end{figure}

We observe that the price spread exhibits properties of a stationary series. An Augmented Dickey-Fuller (ADF) test conducted on the HEFTCom2024 offline data yields a p-value as low as 1e-29, rejecting the hypothesis of non-stationarity. Nevertheless, the price spread still shows signs of weak seasonality. Figure \ref{fig5} illustrates the mean and variability of the price spread at different hour-of-day. The statistics are conducted across different periods from 2022-01 to 2024-01, with each period spanning half a year.

Across multiple statistical periods, the price spread exhibits varying mean values at different hour-of-day. This indicates that the price spread is not entirely random with respect to the hour-of-day. Another key observation is that the daily volatility pattern of the price spread remains consistent across different half-year testing periods. Specifically, the price spread curve throughout the day shows a bimodal pattern, with lower values from early morning to noon and typically peaking at 5:00 and 15:00. It can be inferred that such patterns are related to the daily patterns of the solar generation and load demand in the market. However, since the real-time price component of the price spread is determined by the system imbalance, which is an unobservable and stochastic factor, the conditional distribution of the price spread is characterized by high variability at each hour. As a result, the effectiveness of any time series forecasting models for the price spread is inherently limited. Accordingly, focusing on the mean behavior of the spread may offer a more robust and informative modeling direction.

\subsection{Stochastic Trading Model}
Based on the above analysis, we propose a stochastic trading (ST) model to optimize trading revenue based on the statistical features of the price spread. The proposed strategy can be formulated as an optimization problem under price spread uncertainty, expressed as
\begin{equation}
    \begin{aligned}
    \max_{\hat{e}_1,\dots,\hat{e}_{48}} \mathbb{E}[S_{trading}]=\mathbb{E}[\sum_{t=1}^{48}\pi_{\text{D},t}\hat{e}_t-0.07(\hat{y}_t-\hat{e}_t)^2]+ \sum_{t=1}^{48} \pi_{\text{SS},t}y_t, \\
    s.t \quad 0 \leq \hat{e}_t \leq 1800 \quad \forall t=1,2,...,48,
    \end{aligned}
\end{equation}
where the bidding quantity is constrained by the competition rules to be within the range from 0 MWh to 1800 MWh. Point forecasts $\hat{y}_t$ are used for power generation. Assuming that the real power generation $y_t$ and $\pi_{\text{D}}$ are decoupled, the above problem can be transformed into
\begin{equation}
    \begin{aligned}
        \label{SP}
        \max_{\hat{e}_0,\dots,\hat{e}_{48}} \sum_{t=1}^{48}\mathbb{E}(\pi_{\text{D},t}|t)\hat{e}_t-0.07(\hat{y}_t-\hat{e}_t)^2.
    \end{aligned}
\end{equation}

As analyzed in Section \ref{sec41}, if the expected price spread over a period of time can be considered approximately stationary conditioned on $t$, then the mean of the price spread from the bidding time to the historical data within a certain time window, $\hat{\mathbb{E}}(\pi_{\text{D},t}|t)$, can be used to estimate the expected price spread, $\mathbb{E}(\pi_{\text{D},t}|t)$. During the online phase, $\hat{\mathbb{E}}(\pi_{\text{D},t}|t)$ can be updated iteratively using the latest price spread data, formulated as
\begin{equation}
    \begin{aligned}
        \hat{\mathbb{E}}(\pi_{\text{D},t}|t)=\frac{1}{\left\lfloor w/48 \right\rfloor}\sum_{i=1}^{w}\pi_{\text{D},i}\mathbb{I}(i \bmod48=t),
    \end{aligned}
\end{equation}
where $w$ denotes the length of the time window, and $\left\lfloor w/48 \right\rfloor$ is the number of complete days in the time window. After offline tuning and testing, the length of the time window is set to two months ($w=2880$) before the bidding time in the final solution, as it provides acceptable performance. 

Furthermore, since the price spread lacks strong sequential features, let $r(\hat{e}_t)$ denote the revenue at time $t$, Eq. \eqref{SP} can be simplified into an optimization problem at each time step $t$, expressed as
\begin{equation}
    \begin{aligned}
        \label{SP2}
    \max_{\hat{e}_t \in [0,1800]} r(\hat{e}_t)= \hat{\mathbb{E}}(\pi_{\text{D},t}|t)\hat{e}_t-0.07(\hat{y}_t-\hat{e}_t)^2.
    \end{aligned}
\end{equation}

Given the power forecast $\hat{y}_t$ and the conditional expectation of the price spread $\hat{\mathbb{E}}(\pi_{\text{D},t}|t)$, the optimal bidding decision $\hat{e}^{*}_t$ can be obtained through the Karush-Kuhn-Tucker (KKT) conditions of Eq. \eqref{SP2}. The optimal bidding strategy is given by
\begin{equation}
    \label{optimal}
    e^{*}_t (\hat{y}_t,\hat{\mathbb{E}}(\pi_{\text{D},t}|t)) =\begin{cases}
    1800, \quad \text{if } \hat{y}_t+7.14\hat{\mathbb{E}}(\pi_{\text{D},t}|t) \geq 1800, \\
    0, \quad \text{if } \hat{y}_t+7.14\hat{\mathbb{E}}(\pi_{\text{D},t}|t) < 0, \\
    \hat{y}_t+7.14\hat{\mathbb{E}}(\pi_{\text{D},t}|t),\quad \text{otherwise}.
    \end{cases}
\end{equation}

\subsection{Power Forecasting Model for Trading} 
\label{trading analysis}
Based on the ST model above, the impact of power and price spread forecast errors on the trading revenue can be quantified. This analysis can help design a suitable power forecasting model for trading tasks.

The theoretical optimal bidding can be represented as $e^{*}_t (y_t,\mathbb{E}(\pi_{\text{D},t}|t))$. The gap between the actual revenue caused by the forecasts and the theoretically optimal revenue is defined as the trading loss $\mathcal{L} (\cdot)$, which can be expressed as
\begin{equation}
    \label{SPO loss}
    \mathcal{L}(\hat{y}_t,\hat{\pi}_{\text{D},t},y_t,\pi_{\text{D},t})=r[e^{*}_t (y_t,\mathbb{E}(\pi_{\text{D},t}|t))]-r[e^{*}_t (\hat{y}_t,\hat{\mathbb{E}}(\pi_{\text{D},t}|t))].
\end{equation}

Substituting Eq. \eqref{optimal} into Eq. \eqref{SPO loss}, the relationship between trading loss and forecast errors can be derived as
\begin{equation}
    \begin{aligned}
        \label{trading loss}
    \mathcal{L}(\hat{y}_t,\hat{\pi}_{\text{D},t},y_t,\pi_{\text{D},t}) = 0.07(\hat{y}_t-y_t)^2
    +3.57(\hat{\mathbb{E}}(\pi_{\text{D},t}|t)-\pi_{\text{D},t})^2
    +(\hat{y}_t-y_t)(\hat{\mathbb{E}}(\pi_{\text{D},t}|t)-\pi_{\text{D},t}).
    \end{aligned}
\end{equation}

The trading loss indicates that if the expected estimate of the price spread is accurate, i.e., the expected error is 0, the impact of the power forecast deviation on revenue is dominated by the quadratic term. This suggests that the information about total power generation required for the trading strategy should be more sensitive to quadratic deviations rather than the linear deviations emphasized by quantile regression methods. Therefore, to aligns the forecasting model with the needs of the trading task by learning from trading-related loss, we design an additional power forecasting model for the trading task, utilizing mean squared error (MSE) as the loss function for trading. The main framework of this power forecasting model are similar to that in the forecasting track, but the loss function on the training set is changed from pinball loss to MSE. In other words, the forecast $\hat{y}_t$ required in Eq. \eqref{optimal} is generated by the model $ F'(\mathbf{x}_t;\theta)$, where $\theta$ is optimized by minimizing the MSE loss, expressed as
\begin{equation}
    \begin{aligned}
        \label{MSE}
        \theta^* = \underset{\theta}{\text{argmin}} \quad \frac{1}{|\mathcal{W}_{\text{DWD}}|}\sum_{t=1}^{|\mathcal{W}_{\text{DWD}}|}(F'(\mathbf{x}_t;\theta)-y_t)^2.
    \end{aligned}
\end{equation}

\section{Case Study}
\label{sec5}
\subsection{Evaluation Environments and Metrics}
\label{sec51}
To validate the effectiveness of the proposed hybrid strategy, we design two case studies with different test periods. In both cases, the FLT of the NWP data need to be greater than 23 hours and the reference time need to be at 00:00. The first case includes test data from February 1, 2023, to August 1, 2023, while the remaining data is reserved for training. The second case is designed to replicate the competition conditions, using a test period from February 20, 2024, to May 19, 2024, and the remaining data for training. In the second case, we have identified missing data from the GFS model and supplemented it with corresponding DWD data. To comprehensively evaluate the performance of probabilistic forecasts, we adopt the mean pinball loss (MPL), mean continuous ranked probability score (MCRPS), mean Winkler Score (MWS) as evaluation metrics. The MPL is defined in Eq. \eqref{eq:MPL}. The CRPS evaluates the accuracy of the forecast while also considering the rationality of the CDF. The definition of MCRPS is as
\begin{equation}
    \label{eq:MCRPS}
    \begin{aligned}
        \mathrm{MCRPS} = \frac{1}{N}\sum_{i=1}^{N}\int_{-\infty}^{\infty}(\hat{F}_i(y)-\mathbb{I}(y\geq y_i))^2dy,
    \end{aligned}
\end{equation}
where $\hat{F}_i(y)$ is the CDF of the probabilistic forecast, and $y_i$ is the actual power generation. Winkler Score is a metric that evaluates the accuracy and width of the forecast. It penalizes both the accuracy and width of the forecast, defined as
\begin{equation}
    \label{eq:MWS}
    \mathrm{WS}(\hat{L}, \hat{U}, y ; \alpha)=\left\{\begin{array}{ll}
        \hat{U}-\hat{L}+\frac{2}{\alpha}(\hat{L}-y), & \text { if } y<\hat{L} \\
        \hat{U}-\hat{L}, & \text { if } \hat{L} \leq y \leq \hat{U} \\
        \hat{U}-\hat{L}+\frac{2}{\alpha}(y-\hat{U}), & \text { if } y>\hat{U}
        \end{array}\right.
\end{equation}
where $\hat{L}$ and $\hat{U}$ are the lower and upper bounds of the prediction interval corresponding to the given confidence level $1-\alpha$. In the subsequent experiments, the 10\% and 90\% quantile forecasts are used as the lower and upper bounds, respectively, corresponding to $\alpha=0.2$ and a confidence level of 80\%. The MWS on the test set is defined as
\begin{equation}
    \label{eq:MWS_test}
    \begin{aligned}
       \mathrm{MWS} = \frac{1}{N}\sum_{i=1}^{N}\mathrm{WS}(\hat{L}_i, \hat{U}_i, y_i ; \alpha),
    \end{aligned}
\end{equation}
where $\hat{L}_i$ and $\hat{U}_i$ denote the lower and upper bounds of the prediction interval for the $i$-th sample.

\subsection{Evaluation of Stacking Sister Models}
We compare the performance of single NWP models and stacked sister models that combine multiple NWP sources in the two cases. The results are shown in Table \ref{tab:performance_sis}, where the wind power and solar power forecasts are evaluated separately.

\begin{table}[!htbp]
    \centering
    \caption{Performance comparison of single NWP and stacked sister models for wind and solar power forecasting}
    \setlength{\tabcolsep}{3pt}
    \renewcommand{\arraystretch}{1.1} 
    \begin{tabular}{l@{\hskip 4pt}ccc@{\hskip 6pt}ccc@{\hskip 6pt}ccc@{\hskip 6pt}ccc}
        \toprule
        \multirow{2}{*}{Approaches} & \multicolumn{3}{c}{Case I (Wind)} & \multicolumn{3}{c}{Case II (Wind)} & \multicolumn{3}{c}{Case I (Solar)} & \multicolumn{3}{c}{Case II (Solar)} \\
        \cmidrule(lr){2-4} \cmidrule(lr){5-7} \cmidrule(lr){8-10} \cmidrule(lr){11-13}
        & MPL & MCRPS & MWS & MPL & MCRPS & MWS & MPL & MCRPS & MWS & MPL & MCRPS & MWS \\
        \midrule
        DWD      &   28.96   &   53.18   &   357.72   &   18.90   &  34.66    &   230.98   &   13.35   &  24.25    &   164.28   &   13.72   &   24.91   &  157.30   \\
        GFS      &   30.44   &   55.87   &   381.51   &   19.19   &  35.27    &   243.49   &   17.54   &  32.01    &   218.20   &   15.07   &   27.35   &  171.54  \\
        Stacking &   27.13   &   49.74   &   334.65   &   17.69   &  32.46    &   220.20   &   13.33   &  24.25    &   163.84   &   13.68   &   24.86   &  156.06   \\
        \bottomrule
    \end{tabular}
    \label{tab:performance_sis}
\end{table}
In the wind power forecasting task, the stacked models significantly reduce the MPL, MCRPS, and MWS compared to models trained on a single NWP in both cases. For the MPL metric, which is the focus of the competition, the stacked models achieve improvements of 10.87\% and 6.32\% compared to the GFS and DWD models in Case I. In Case II, the stacked models achieve improvements of 7.82\% and 6.40\% compared to the GFS and DWD models, respectively.
However, for the solar power forecasting task, the effect of the stacked models is minimal, only reducing the MPL by 0.15\% and 0.29\% in Case I and Case II, respectively, compared to using only the NWP provided by DWD. Even in Case I, the MCRPS of the stacked models is slightly higher than that of the models trained on a single NWP. The detailed reason for this result is explained in Section \ref{sec6}.

\subsection{Evaluation of Online Post-processing}
To simulate rolling post-processing, we use 10 days of data from February 20, 2024 to March 2, 2024 as the baseline data without post-processing. From March 3 onwards, a rolling test is performed daily. Specifically, data from February 20 to day \textit{n} is used as the training set to predict solar power for day \textit{n}+2, as the actual available solar power typically lags one day. The offline model without post-processing and the offline post-processing model, which only uses the first 10 days of data to train the parameters of the post-processing model, are used as baseline comparisons. The comparative results are shown in Table \ref{tab:performance_online}.

\begin{figure*}[!ht]
\centering
\includegraphics[width=1\textwidth]{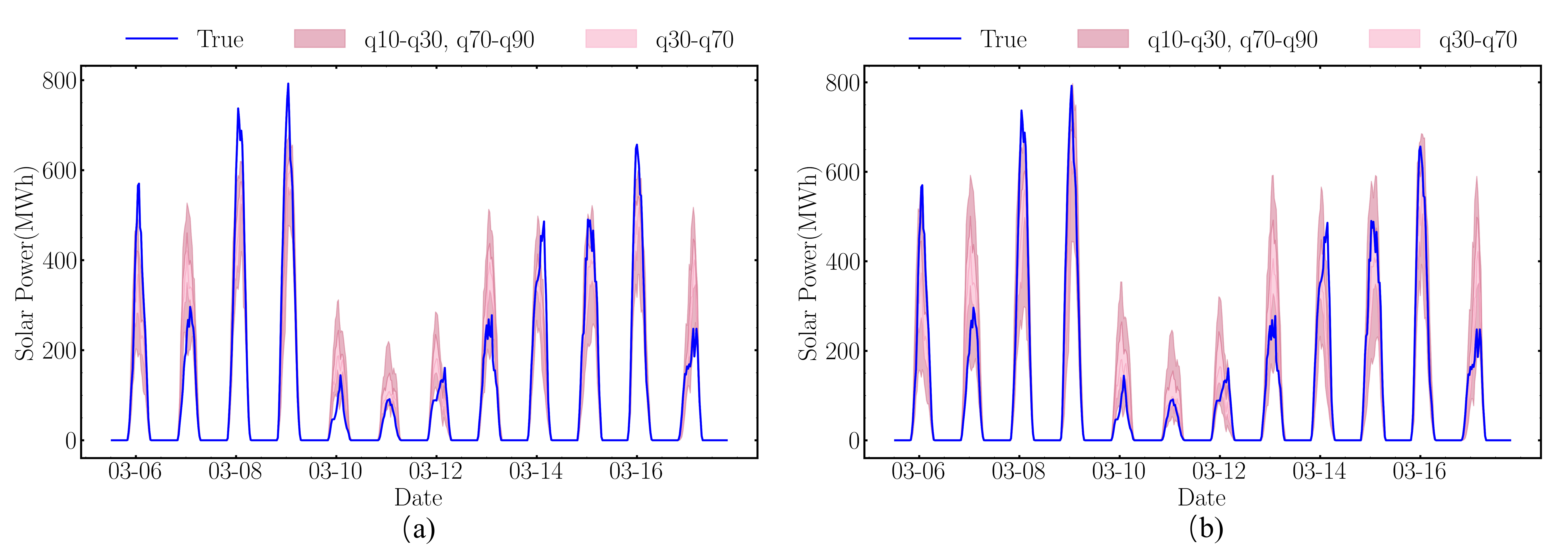}
\caption{Sub-figure (a): Actual solar power values and quantile forecasts without post-processing. Sub-figure (b): Actual solar power values and quantile forecasts with online post-processing. Different colored blocks represent different quantile forecast results. The legend “q10” represents the forecast result at the 10\% quantile, and so on.}
\label{fig3}
\end{figure*}

\begin{table}[htbp]
    \centering
    \caption{Performance comparison of original and post-processed forecasts}
    \setlength{\tabcolsep}{2pt} 
    \renewcommand{\arraystretch}{1.2}
    \begin{tabular}{cccc}
      \toprule
      Approaches & MPL & MCRPS & MWS \\
      \midrule
      Offline Model & 15.22 & 27.18 & 188.16 \\
      Offline Post-processing & 14.22 & 25.85 & 161.00 \\
      Online Post-processing  & 13.72 & 24.92 & 157.30 \\
      \bottomrule
    \end{tabular}
    \label{tab:performance_online}
  \end{table}
The rolling test results show that compared to the model trained solely on the offline dataset, the MPL, MCRPS, and MWS are reduced by 9.86\%, 8.31\%, and 16.4\%, respectively, and compared to the offline post-processing model, the MPL, MCRPS, and MWS are also reduced by 3.52\%, 3.60\%, and 2.3\%, respectively. This also indicates that the online post-processing approach is more capable of dynamically adapting to the changed distribution of solar power generation. Figure \ref{fig3} illustrates the quantile forecast results alongside the actual solar power values. Despite the reduced sharpness on certain days, notably, the original forecasts failed to capture the peak solar power values on certain days, (e.g., 03-06, 03-08, 03-09). The proposed online rolling post-processing model overcomes this limitation and achieves better results in MCRPS and MWS.

\subsection{Evaluation of Aggregated Quantile Forecasting}
To evaluate the effectiveness of the proposed quantile aggregation method, we compare it against two alternative forecasting strategies: (1) forecasts obtained by summing the forecast results quantile by quantile; and (2) direct total power forecasting using separate models trained on DWD and GFS data, with the predictions combined using the stacking method. To ensure a fair comparison, the hyperparameters of the total power forecasting model are also optimized. However, a major limitation of the direct forecasting approach is the lack of visibility into the individual contributions from wind and solar components. As a result, it is not possible to apply post-processing to each component separately. This is particularly problematic in the online testing phase, where solar capacity expansion and planed outage of Hornsea 1 require dynamic post-processing for adjustment. Therefore, the direct total power forecasting method is only compared in Case I. The results of this comparison are presented in Table \ref{tab:performance_total}.

\begin{table}[htbp]
    \centering
    \caption{Performance comparison of different total power forecasting methods}
    \setlength{\tabcolsep}{4pt}
      \begin{tabular}{lccc ccc}
      \toprule
      \multicolumn{1}{c}{\multirow{2}[0]{*}{Approaches}} & \multicolumn{3}{c}{Case I} & \multicolumn{3}{c}{Case II} \\
      \cmidrule(lr){2-4} \cmidrule(lr){5-7}
            & MPL   & MCRPS & MWS   & MPL   & MCRPS & MWS \\
      \midrule
      Quantile by Quantile & 34.40 & 47.99 & 417.37 & 24.50 & 37.77 & 292.46 \\
      Total Forecasting & 34.86 & 43.41 & 429.19 &    none   &   none    &  none \\
      Aggregation & 34.35 & 46.88 & 418.13 & 24.32 & 36.93 & 288.33 \\
      Team GEB & none & none & none & 25.16 & none & none \\
      \bottomrule
      \end{tabular}
    \label{tab:performance_total}
\end{table}
For the MPL metric, the proposed quantile aggregation method reduces the MPL compared to the summation method quantile by quantile in both cases, with reductions of 0.15\% and 0.73\%, respectively. Compared to directly forecasting the total generation, the proposed method also achieves superior performance with a 1.46\% reduction. For the MCRPS metric, the proposed method shows a more significant reduction compared to the MPL metric, with reductions of 2.31\% and 2.22\%, respectively. This reflects the predicted CDF from the proposed method is more rational than the quantile-by-quantile method. Although the direct total generation forecasting method achieves best fit to the actual CDF (i.e., the lowest MCRPS), it does not achieve better performance in the MPL metric, which is the focus of the competition. The increased input dimensionality required for modeling the aggregated output may lead to a greater risk of overfitting and makes it more challenging to design specialized hyperparameters for individual quantile forecasting models. The recorded performance of the GEB team is also shown in Table \ref{tab:performance_total}, which reflects the actual scores obtained during the HEFTCom2024 competition. It is observed that the proposed method slightly outperforms the GEB team during the competition. This is because our method was gradually developed during the competition, while the case study used the final solution.

\begin{figure*}[!ht]
\centering
\includegraphics[width=1\textwidth]{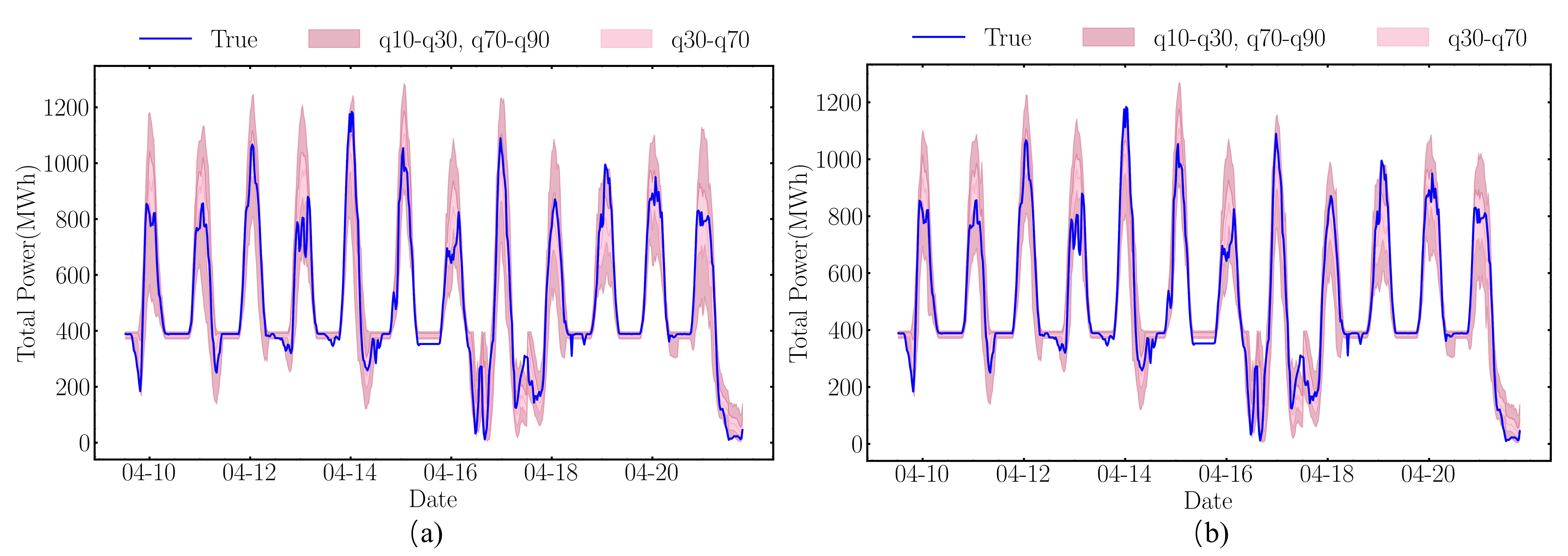}
\caption{Sub-figure (a): Actual total power generation and quantile forecasts produced by adding the wind and solar forecast results quantile by quantile. Sub-figure (b): Actual total power generation and quantile forecasts produced by the proposed quantile aggregation method. Different colored blocks represent different quantile forecast results. The legend “q10” represents the forecast result at the 10\% quantile, and so on.}
\label{fig4}
\end{figure*}

Figure \ref{fig4} illustrates the quantile forecasts and actual total generation for certain periods in Case II. It is observed that the proposed aggregation method primarily reduces the pinball loss at the edge quantiles by narrowing the forecast intervals, when the intervals are wide enough to cover the actual values.

\subsection{Evaluation in Trading Tasks}

An ablation study is conducted to evaluate the effectiveness of the proposed ST strategy. The total revenue of series trading strategies is compared in Case I and Case II, which are listed as follows:
\begin{itemize}
    \item Bid as power forecasts at 50\% quantile: The benchmark strategy provided by HEFTCom2024.
    \item Bid as MSE-oriented power forecasts, which is detailed in Section \ref{trading analysis}.
    \item Naive Forecasting: The power forecast is provided by the MSE-oriented model, and the price spread is predicted by the mean of the price spread in the past 2 months, which is denoted as $\hat{\pi}_{\text{D},t}=\bar{\pi}_{\text{D}}$. The optimal bidding is calculated by Eq. \eqref{optimal} using the predicted price spread and power generation.
    \item Seasonal Persistence Model: Similar to the naive forecasting, but the price spread is predicted by the price spread at the same season on the previous day, which is denoted as $\hat{\pi}_{\text{D},t}=\hat{\pi}_{\text{D},t-48}$. 
    \item AR, ETS, ARMA, SARMAX: Similar to the naive forecasting, but the price spread is predicted by Auto Regressive (AR), Exponential Smoothing (ETS), Auto-Regressive Moving Average (ARMA), and Seasonal Auto-Regressive Moving-Average with Exogenous Regressors (SARMAX), respectively.
    \item ST with q50 power forecasts and ST with MSE-oriented power forecasts (the final solution in the trading track).
\end{itemize}

\begin{table}[htbp]
    \centering
    \caption{Result of the ablation study of trading strategies on revenue}
    \setlength{\tabcolsep}{3pt} 
    \renewcommand{\arraystretch}{1.2}
    \begin{tabular}{lcccc}
      \toprule
      \textbf{Approaches} & \textbf{Case I (£m)} & \textbf{$\Delta$ (\%)} & \textbf{Case II (£m)} & \textbf{$\Delta$ (\%)} \\
      \midrule
        Bid as q50              & 305.94 & 0.00 & 87.62 & 0.00 \\
        Bid as MSE-oriented     & 306.32 & 0.12 & 87.68 & 0.07 \\
        Naive Forecasting       & 305.61 & –0.11 & 87.56 & –0.07 \\
        Seasonal Persistence Model & 306.10 & 0.05 & 87.68 & 0.07 \\
        AR                      & 306.24 & 0.10 & 87.67 & 0.06 \\
        ETS                     & 306.43 & 0.16 & 87.81 & 0.22 \\
        ARMA                    & 306.25 & 0.10 & 87.67 & 0.06 \\
        SARMAX                  & 306.26 & 0.10 & 87.68 & 0.07 \\
        ST (q50)                & 307.29 & 0.44 & 88.02 & 0.46 \\
        ST (MSE-oriented)       & 307.39 & 0.47 & 88.05 & 0.49 \\
      \bottomrule
    \end{tabular}
    \label{tab:performance_trading}
\end{table}

Table \ref{tab:performance_trading} summarizes the revenue performance of various trading strategies across two cases. Although a variety of spread forecasting methods are compared to provide informative signals for trading, the proposed trading strategy consistently achieved the best overall performance in both cases, with the improvement of roughly 0.47\%-0.49\% over the “bid as forecast” baselines. Besides, the table reveals several noteworthy observations.

For the power generation forecast required by the trading strategy, Simply changing the loss function of the power forecasting model from the pinball loss at 50\% quantile to MSE can achieve a modest revenue improvement (e.g., from £305.94 m to £306.32 m in Case I), as MSE is more similar to the downstream trading loss structure. 

For the naive forecasting approach, while the analyzed price spreads exhibit clear stationarity, this method fails to account for the seasonal variation, particularly the hour-of-day effect. Although the seasonal persistence model captures the seasonal pattern of price spreads, the spreads are subject to significant randomness and volatility due to underlying system imbalances. Even after accounting for the hour-of-day as covariate information, the variance of the spreads remains substantial. As a result, simply shifting the previous day's spreads forward fails to produce a smooth forecast.

In the case of autoregressive or smoothing models such as AR, ETS, ARMA, and SARMAX, despite careful hyperparameter tuning and the use of several variants, these models remain sensitive to noise and prone to overfitting to high-frequency fluctuations-none outperform the baseline by more than a million pounds. Compared to AR and ARMA, the performance of ETS and SARMAX models, which incorporate seasonality, shows only slight improvement. In contrast, the proposed ST method can be viewed as a persistence model conditioned on the expected value of hour-of-day, focusing on capturing the long-term trend of the spread rather than the short-term noise-induced fluctuations that are difficult to estimate.

\section{Discussion}
\label{sec6}
\subsection{Tricks and Lessons}
In general, we believe that identifying the key issues in the HEFTCom2024 forecasting and trading tasks and designing tailored methods is crucial to achieving high performance. We tend to choose a combination of multiple lightweight and well-established methods to address these specific problems. This allows for developing each module independently. For the first issue, results of case studies show a reduction in pinball loss when combining multi-source NWP data in wind power forecasting. The stacking technique provides a flexible and robust ensemble method. However, the effectiveness of stacking hinges on combining strong yet diverse base learners. In our view, the discrepancies in forecast performance among sister models are largely attributable to variations in the spatial coverage of the NWP data. As illustrated in Section 2.3, for the Hornsea 1 area, the NWP coordinates of DWD and GFS do not completely overlap. Although predictions solely from GFS underperform compared to DWD, their combination via stacking actually yields superior results. Conversely, in solar power forecasting, the NWP data share identical coverage areas, meaning that the stacking essentially combines only the underlying NWP model differences, which likely leads to a minimal impact. For the second issue, the change in installed capacity results in a new distribution of solar data. Directly correcting the power forecasts based on the proportion of capacity changes is unreliable, as the growth in capacity is not spatially uniform, and the capacity data is not perfect. It is more reliable to address the distribution shift through data-driven methods, such as learning an online post-processing model. This method can also be applied to other online forecasting scenarios with a deterministic and simple distribution shift. Third, for tasks that require predicting aggregate totals, a decomposition-based forecast is generally appropriate, as it allows for targeted handling of individual components. The proposed method offers a theoretically seamless trick for summing individual probabilistic forecasts and is adaptable to various models capable of learning the pinball loss, although it is computationally expensive as a sufficient number of models need to be trained to fit the CDF. However, given that sufficient offline preparation time is available, the computational cost associated with enhancing accuracy is of relatively minor concern. In our final solution, employing 101 quantile levels is found to achieve an acceptable trade-off between predictive resolution of CDF and training efficiency, with the complete training process requiring approximately 4-5 hours. While increasing the number of quantiles enables a finer reconstruction of the CDF, however, due to the multiplicative nature of the model configurations involved, the resulting increase in training time may hinder overall development efficiency. Besides, due to different forecasting tasks and resource constraints, it is advisable to conduct a case-by-case analysis to determine the most suitable number of quantile models. Fourth, although the price spread is a stationary series with low signal-to-noise ratio, its seasonal characteristics still provide valuable information for enhancing trading performance. The proposed ST method can achieve higher revenue compared to “bid as power forecast” and AR, MA or persistence based models, as it is less sensitive to noise.

Furthermore, frequent retrospective analysis and real-time data management are critical in this online competitions. This habit helped us detect solar distribution shifts and train post-processing models using the collected data. We also emphasize the importance of maintaining strict offline-online testing alignment, specifically by creating test sets that closely resemble the online testing environment to validate the developed models. This approach helps to avoid the impact of bugs on performance.

\subsection{Further Improvements to Trading Strategies} 
The trading loss that is influenced by both power forecast errors and price spread forecast errors is derived Section \ref{trading analysis}. However, the ST strategies in the final solution are not designed to minimize the downstream trading loss in an end-to-end way. According to the trading loss illustrated in Eq.\eqref{trading loss}, improving the accuracy of each forecast is not the only key factor to high revenue. Therefore, we explore the method to further improve trading revenue.

\begin{figure}[!ht]
\centering
\includegraphics[width=0.45\textwidth]{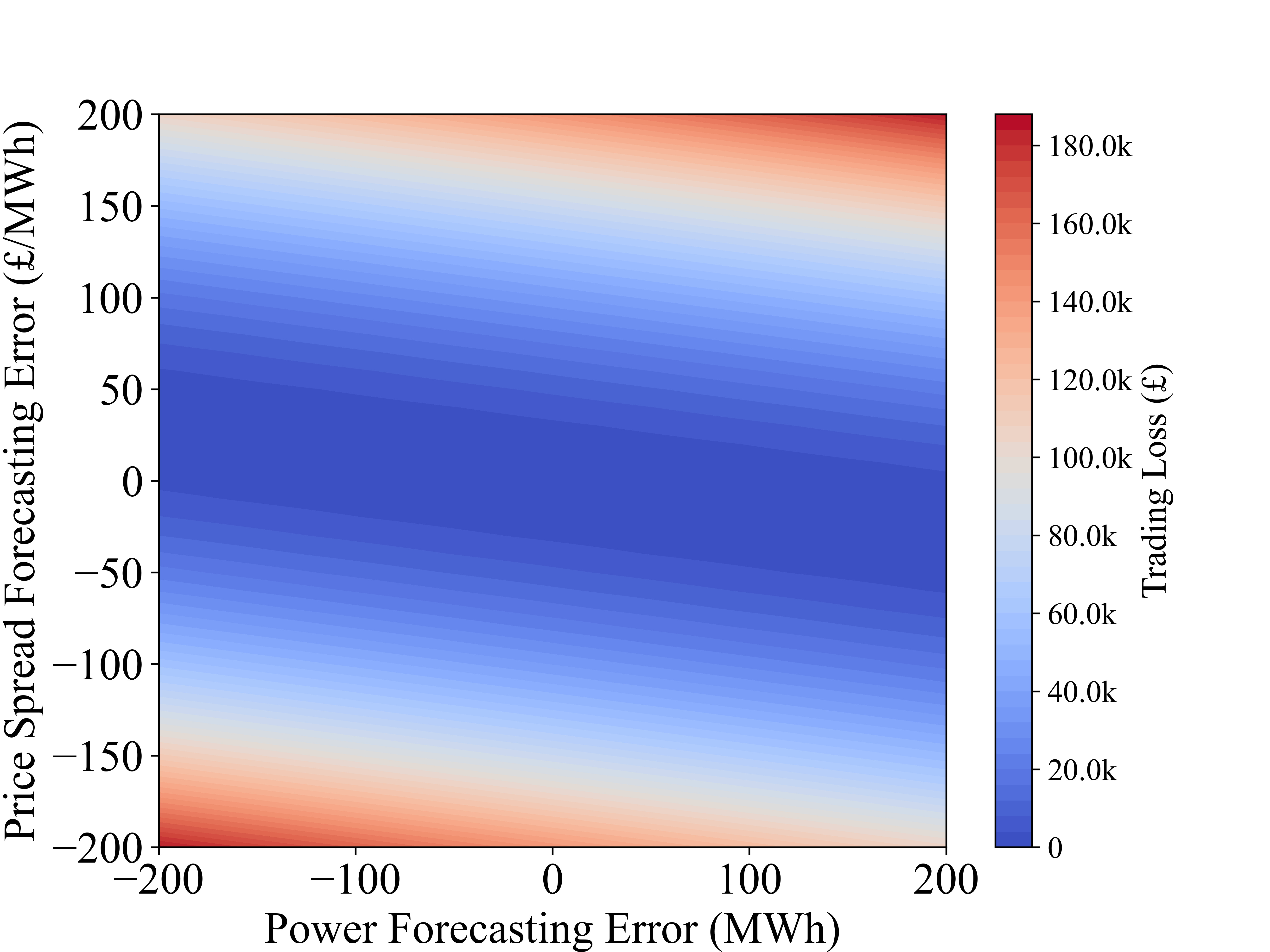}
\caption{Heatmap of trading loss with respect to the forecast errors of power generation and price spread. The color represents the magnitude of the trading loss.}
\label{loss_curve}
\end{figure}

The heatmap of trading loss with respect to the power forecast error $\hat{y}-y$ and price spread forecast error $\hat{\pi}_{\text{D}}-\pi_{\text{D}}$ is shown in Figure \ref{loss_curve}. It reveals that the impact of power forecast errors and price forecast errors on trading loss is asymmetric. For instance, with the same level of power forecast error, trading losses differ depending on whether the price forecast error is positive or negative.

To utilize this asymmetry, an effective strategy is to train an end-to-end trading model that learns the downstream trading loss directly. We provided a formulation of the E2E model for this trading task, as:
\begin{subequations} \label{eq:example}
    \begin{align}
        \underset{\theta_p,\theta_{\pi}}{\text{min}} \quad & \mathbb{E}[\mathcal{L}(\hat{y}(\theta_p),y,\hat{\pi}_{\text{D}}(\theta_{\pi}),\pi_{\text{D}})], \label{eq:examplea} \\
        s.t. \quad & \hat{\pi}_{\text{D}}(\theta_{\pi}) = f(\mathbf{x}_{\pi};\theta_{\pi}) \label{eq:exampleb}, \\
        & \hat{y}(\theta_p) = g(\mathbf{x}_{p};\theta_{p}),  \\
        & \| \hat{y}(\theta_p)-y \|_2^2 \leq \epsilon_p \label{eq:examplec}, \\
        & \| \hat{\pi}_{\text{D}}(\theta_{\pi})-\pi_{\text{D}} \|_2^2 \leq \epsilon_{\pi}. \label{eq:exampled}
    \end{align}
\end{subequations}
where $f(\cdot)$ models the price spread and $g(\cdot)$ models the power generation, and $\theta_p$ and $\theta_{\pi}$ are the parameters of the models, respectively. Parameters $\mathbf{x}_{\pi}$ and $\mathbf{x}_{p}$ are the input features for the price spread and power forecasting models, respectively. Eq.\eqref{eq:examplec} and \eqref{eq:exampled} are set to ensure the accuracy of the forecasts while optimizing the downstream objective. $\epsilon_p$ and $\epsilon_{\pi}$ are the tolerances for the power and price spread forecast errors, respectively. 

To validate the effectiveness of the proposed improvement, we conducted detailed experiments on the Case I and Case II datasets. A multilayer perceptron (MLP) is employed as the E2E trading model, as it allows for flexible loss function design and optimizing the trading loss through differentiable backpropagation. The input features consist of the total power forecast, one-hot encoded hour-of-day. Additionally, an extra benchmark is introduced, which disregards the surface shape of trading loss with respect to the two error components, denoted as “accuracy-oriented” in Table \ref{tab:performance_e2e}. This benchmark shares identical input features with the proposed E2E trading model but adopts MSE as the loss function to predict the price spread. The optimal bidding strategy is derived based on the KKT conditions. The test results on both cases are presented in Table \ref{tab:performance_e2e}. It can be observed that the proposed E2E trading model achieves the highest revenue in both cases. Compared to the “bid as q50” and ST (MSE-oriented) methods, the revenue increases by 0.62\% and 0.15\% in Case I, and by 0.64\% and 0.15\% in Case II, respectively. Although the absolute percentage gains may appear limited, they carry practical significance in view of the low signal-to-noise ratio of the price spread. Notably, the ST-based strategy improved revenue by just 0.49\% over the baseline, but this modest advantage translated into a 3rd-place finish among all teams in HEFTCom2024.

\begin{table}[htbp]
    \centering
    \caption{Performance comparison of E2E trading model and baseline models}
    \setlength{\tabcolsep}{4pt}
      \begin{tabular}{lcc cc}
      \toprule
      \multicolumn{1}{c}{\multirow{2}[0]{*}{Approaches}} & \multicolumn{2}{c}{Case I} & \multicolumn{2}{c}{Case II} \\
      \cmidrule(lr){2-3} \cmidrule(lr){4-5}
            & Revenue (£m)   & Trading Loss (£m)    & Revenue (£m)   & Trading Loss (£m)  \\
      \midrule
      bid as q50 & 305.94 & 82.92 & 87.62 & 16.91 \\
      ST(final solution) & 307.38 & 81.47 & 88.05 &   16.48   \\
      Accuracy-oriented & 306.43 & 82.43 & 87.82 &    16.71    \\
      E2E & 307.84 & 81.02 & 88.18 & 16.35  \\
      \bottomrule
      \end{tabular}
    \label{tab:performance_e2e}
\end{table}

\begin{figure}[!ht]
\centering
\includegraphics[width=1\textwidth]{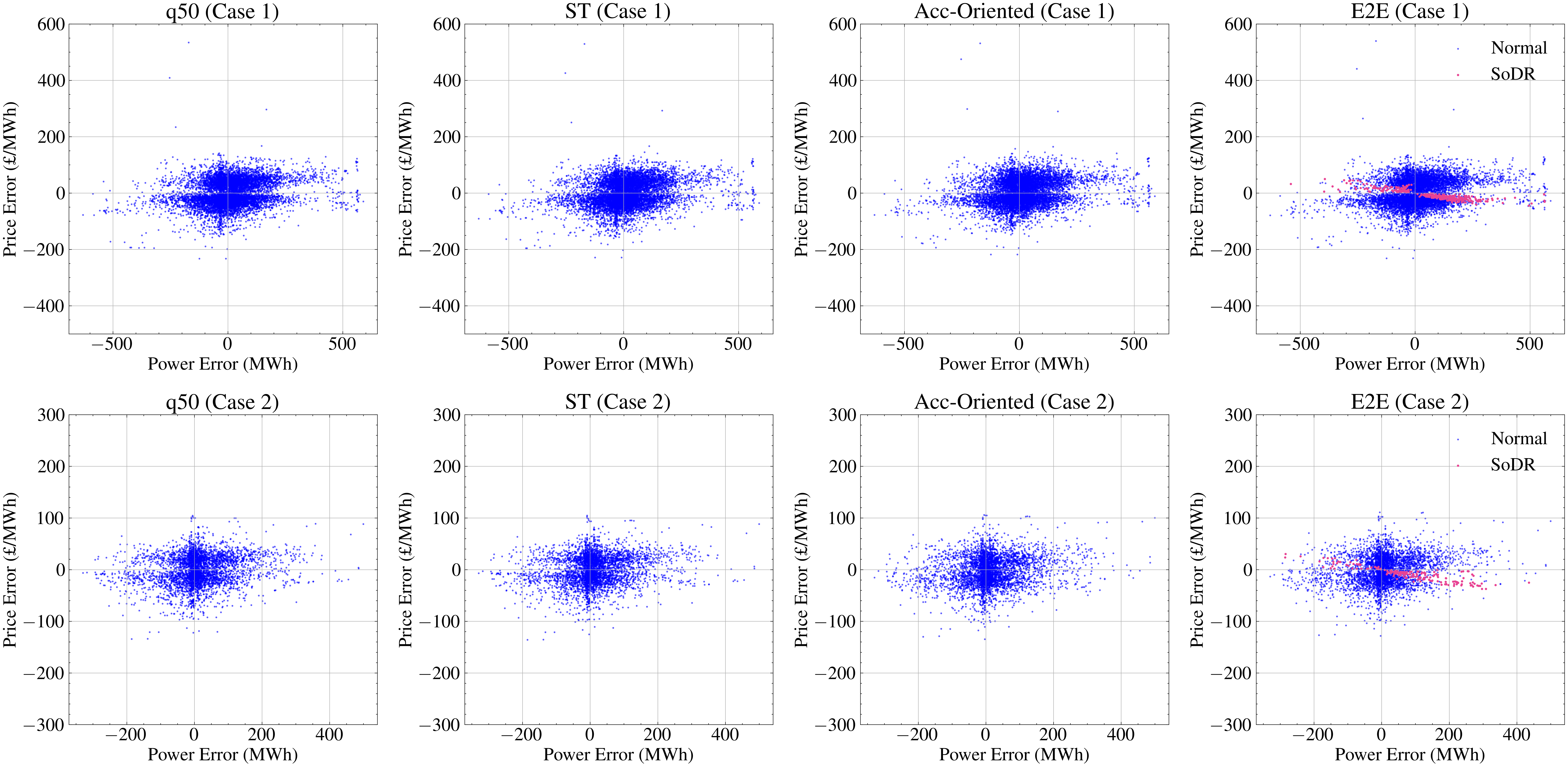}
\caption{Scatter plots of forecast errors in power generation and price spread under different forecasting methods. In the subplot which denotes the E2E approach, the highlighted pink points indicate cases where the method achieves higher trading revenue compared to the ST method despite having larger forecast errors, illustrating the benefit of aligning forecasts with decision-making objectives.}
\label{loss_scatter}
\end{figure}

To further demonstrate how the E2E approach enhances trading performance by shaping the error distribution, we also visualize the power forecast errors and price spread forecast errors of the compared methods in the scatter plot shown in Figure \ref{loss_scatter}. Compared to the bid-as-q50 method (i.e., the estimated price spread is always zero), the ST (final solution of GEB) and accuracy-oriented methods primarily improve trading revenue mainly through more precise estimation of the price spread, which results in an error distribution that is more tightly clustered around zero. Notably, the error distribution of the E2E approach is better aligned with the shape of the trading loss surface depicted in Figure 8. To highlight the differences between the proposed method and others, we emphasize a subset of points in the subplot of E2E approach. These points exhibit higher revenue compared to the ST method but also have higher forecast errors. They tend to be distributed with a certain tilt and are primarily located in regions associated with low trading loss, while appearing less frequently in high-loss regions. Overall, since the trading model is fixed as a deterministic algorithm, the trading revenue is mainly determined by the quality of the forecasts. Even with imperfect predictions, the trading revenue can be improved by training forecasts that are closer to the downstream decision-making loss.

\section{Conclusion}
\label{sec7}
This paper discusses the methodology developed by the GEB team in HEFTCom2024. We summarize the key factor that contributed to our performance, which is a tailored hybrid strategy that includes stacked multi-NWP sister forecasting models for wind power, online post-processing models for solar distribution shift, quantile aggregation for total generation forecasting, and stochastic trading strategies. The effectiveness of these methods is validated in HEFTCom2024 online testing and two case studies that cover different time periods and data distributions. We also provide a future improvements to the downstream trading revenue through end-to-end learning approaches. In summary, this paper contributes a detailed and reproducible solution for probabilistic energy forecasting and trading, which is proven effective in such world-class competition. This work can provide valuable insights for researchers and practitioners in the field of energy forecasting and trading. Further research could explore the seamless integration of probabilistic forecasting with downstream stochastic trading tasks.

\section*{Acknowledgment}
The authors would like to thank the organizers of HEFTCom2024 for providing the data and platform. The authors also thank Prof. Jethro Browell (University of Glasgow) and Dr. Linwei Sang (Tsinghua University) for their helpful suggestions on the manuscript. This work is supported by National Natural Science Foundation of China under Grant No. 52407125, No. U2243243 and National Key R\&D Project (No. 2024YFB4206500)

\section*{Code Availability}
As part of this work, we have published two open-source codebases. The first codebase provides all the details of the methods proposed in this paper and enables the reproduction of all the experiments presented herein, which can be found at \url{https://github.com/BigdogManLuo/HEFTcom24}. A permanent archived version is also available via Zenodo at \url{https://doi.org/10.5281/zenodo.15351009}. The second codebase represents the final solution developed by the GEB team during the competition. This is an engineering-oriented codebase capable of performing online forecasting tasks, and can be found at \url{https://github.com/BigdogManLuo/HEFTcom24/tree/online}.

\section*{Author Contributions}
\textbf{Chuanqing Pu} led the development of the solutions and wrote this paper. \textbf{Feilong Fan} funded the work and provided guidance for this paper. \textbf{Nengling Tai} funded the work and offered overall supervision. \textbf{Songyuan Liu} and \textbf{Jinming Yu} contributed to the development of the solutions. All authors reviewed and approved the final version of the manuscript.

\appendix

\clearpage
\appendix
\renewcommand{\thefigure}{A.\arabic{figure}}
\setcounter{figure}{0} 
\renewcommand{\theequation}{A.\arabic{equation}}
\setcounter{equation}{0}

\section*{Appendix A. Independence Test of Wind and Solar Power}

As shown in Eq. (16), the PDF of total generation is obtained by convolving the PDFs of wind and solar power. The accurate estimation of the PDF of total generation depends on whether wind and solar power are conditionally independent. To verify this conditional independence, we used a regression-based test approach as follows.

Assuming that the observed wind and solar power data follow an additive noise model (ANM), i.e.,
\begin{equation}
    \begin{aligned}
        \label{ANM}
        y_{\text{w}}=f_{\text{w}}(\mathbf{x}_{\text{w}})+\epsilon_{\text{w}}\\
        y_{\text{s}}=f_{\text{s}}(\mathbf{x}_{\text{s}})+\epsilon_{\text{s}}
    \end{aligned}
\end{equation}
where $f_{\text{w}}(\cdot)$ and $f_{\text{s}}(\cdot)$ denote the regression functions of wind and solar power, respectively, $\mathbf{x}_{\text{w}}$ and $\mathbf{x}_{\text{s}}$ represent the input features for wind and solar power, and $\epsilon_{\text{w}}$ and $\epsilon_{\text{s}}$ are the additive noise terms with zero mean.

If $\epsilon_{\text{w}}$ and $\epsilon_{\text{s}}$ are independent, wind and solar power are independent conditioned on $\mathbf{x}_{\text{w}}$ and $\mathbf{x}_{\text{s}}$ respectively. Therefore, we established two separate regression models $\hat{f}_{\text{w}}(\cdot)$ and $\hat{f}_{\text{s}}(\cdot)$ using LightGBM, and the estimation of residual terms extracted from the models are as follows:
\begin{equation}
    \begin{aligned}
        \label{residual}
        \hat{\epsilon}_{\text{w}}=y_{\text{w}}-\hat{f}_{\text{w}}(\mathbf{x}_{\text{w}})\\
        \hat{\epsilon}_{\text{s}}=y_{\text{s}}-\hat{f}_{\text{s}}(\mathbf{x}_{\text{s}})
    \end{aligned}
\end{equation}

We used mutual information as the indicator for the conditional independence test, and the calculation method is as follows:
\begin{equation}
    \begin{aligned}
        \label{MI}
        \text{MI}(\hat{\epsilon}_{\text{w}},\hat{\epsilon}_{\text{s}})=\sum_{\hat{\epsilon}_{\text{w}} \in \mathcal{W}} \sum_{\hat{\epsilon}_{\text{s}} \in \mathcal{S}} p(\hat{\epsilon}_{\text{w}},\hat{\epsilon}_{\text{s}}) \log \frac{p(\hat{\epsilon}_{\text{w}},\hat{\epsilon}_{\text{s}})}{p(\hat{\epsilon}_{\text{w}})p(\hat{\epsilon}_{\text{s}})}
    \end{aligned}
\end{equation}
where $\mathcal{W}$ and $\mathcal{S}$ are the sets of residual terms of wind and solar power, respectively, and $p(\hat{\epsilon}_{\text{w}},\hat{\epsilon}_{\text{s}})$, $p(\hat{\epsilon}_{\text{w}})$, and $p(\hat{\epsilon}_{\text{s}})$ are the joint and marginal probabilities of the residual terms, respectively.

\begin{figure}[!ht]
\centering
\includegraphics[width=0.45\textwidth]{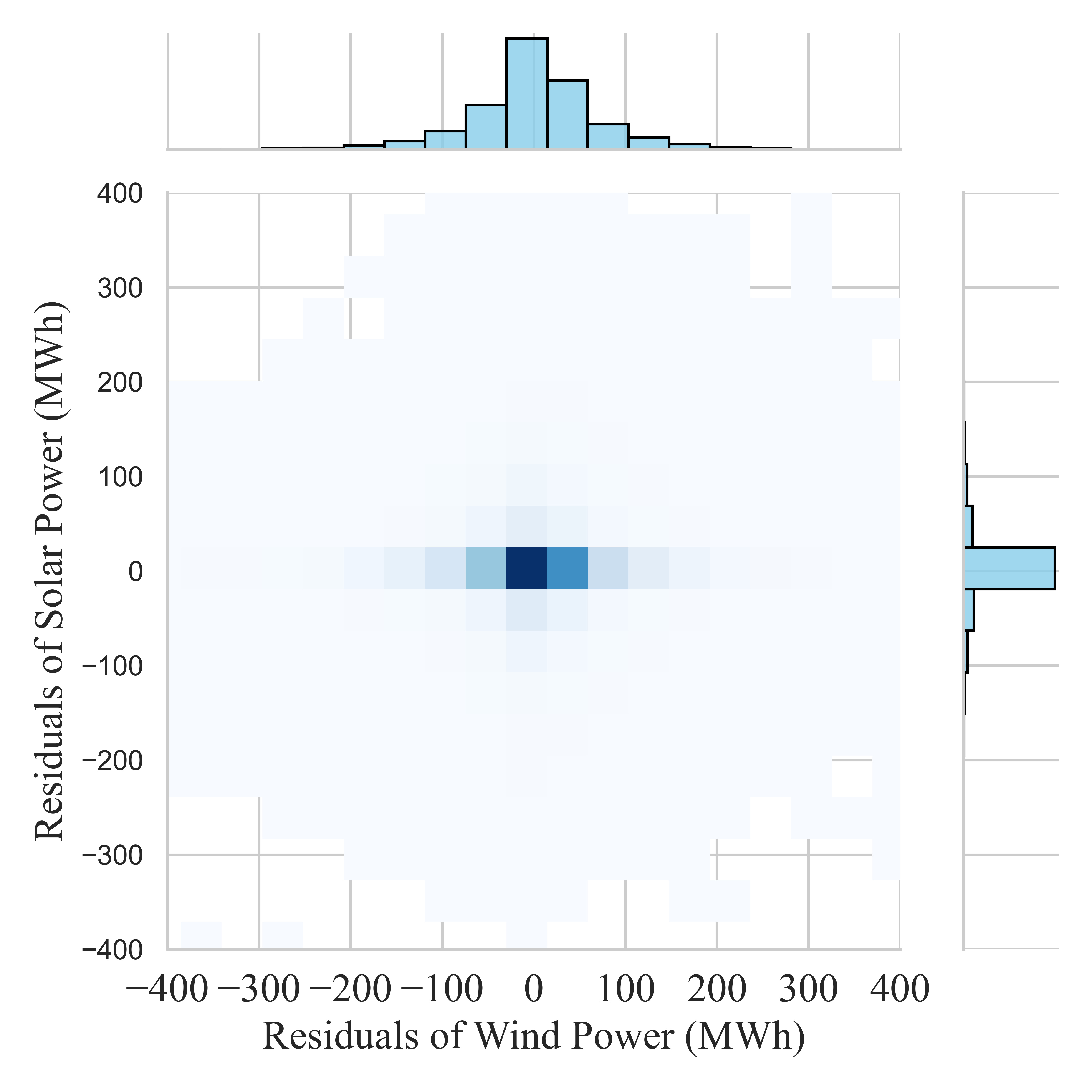}
\caption{The distributions of the residuals of wind and solar power.}
\label{fig:appendix_a1}
\end{figure}

We conducted experiments on the data from September 2020 to January 2024, and the results show that $\text{MI}(\hat{\epsilon}_{\text{w}},\hat{\epsilon}_{\text{s}})$ is 0.006, which indicates that wind and solar power can be considered conditionally independent. The distributions of the two residuals are shown in Figure \ref{fig:appendix_a1}. The means of the two residuals are close to 0, and the joint distribution of the residuals is similar to the independent distribution in each dimension.

\begin{thebibliography}{00}

\bibitem[Petropoulos et al., 2022]{Background1}
Petropoulos, F., Apiletti, D., Assimakopoulos, V., Babai, M. Z., Barrow, D. K., Taieb, S. B., ... \& Ziel, F. (2022). Forecasting: theory and practice. \textit{International Journal of Forecasting}, 38(3), 705-871.

\bibitem[Sweeney et al., 2020]{Background2}
Sweeney, C., Bessa, R. J., Browell, J., \& Pinson, P. (2020). The future of forecasting for renewable energy. Wiley Interdisciplinary Reviews: \textit{Energy and Environment}, 9(2), e365.

\bibitem[Hong et al., 2020]{Background3}
Hong, T., Pinson, P., Wang, Y., Weron, R., Yang, D., \& Zareipour, H. (2020). Energy forecasting: A review and outlook. \textit{IEEE Open Access Journal of Power and Energy}, 7, 376-388.

\bibitem[Roach, 2019]{GEFcom1}
Roach, C. (2019). Reconciled boosted models for GEFCom2017 hierarchical probabilistic load forecasting. \textit{International Journal of Forecasting}, 35(4), 1439-1450.

\bibitem[Smyl, Hua, 2019]{GEFcom2}
Smyl, S., \& Hua, N. G. (2019). Machine learning methods for GEFCom2017 probabilistic load forecasting. \textit{International Journal of Forecasting}, 35(4), 1424-1431.

\bibitem[Nagy et al., 2016]{GEFcom3}
Nagy, G. I., Barta, G., Kazi, S., Borbély, G., \& Simon, G. (2016). GEFCom2014: Probabilistic solar and wind power forecasting using a generalized additive tree ensemble approach. \textit{International Journal of Forecasting}, 32(3), 1087-1093.

\bibitem[Landry et al., 2016]{GEFcom4}
Landry, M., Erlinger, T. P., Patschke, D., \& Varrichio, C. (2016). Probabilistic gradient boosting machines for GEFCom2014 wind forecasting. \textit{International Journal of Forecasting}, 32(3), 1061-1066.

\bibitem[Huang et al., 2016]{GEFcom5}
Huang, J., \& Perry, M. (2016). A semi-empirical approach using gradient boosting and k-nearest neighbors regression for GEFCom2014 probabilistic solar power forecasting. \textit{International Journal of Forecasting}, 32(3), 1081-1086.

\bibitem[Landgraf, 2019]{feature-engi}
Landgraf, A. J. (2019). An ensemble approach to GEFCom2017 probabilistic load forecasting. \textit{International Journal of Forecasting}, 35(4), 1432-1438.

\bibitem[Browell et al., 2020]{hyper-opt}
Browell, J., Gilbert, C., Tawn, R., \& May, L. (2020, September). Quantile combination for the EEM20 wind power forecasting competition. In \textit{2020 17th International Conference on the European Energy Market (EEM)} (pp. 1-6). IEEE.

\bibitem[Ganaie et al., 2022]{stacking_review}
Ganaie, M. A., Hu, M., Malik, A. K., Tanveer, M., \& Suganthan, P. N. (2022). Ensemble deep learning: A review. \textit{Engineering Applications of Artificial Intelligence}, 115, 105151.

\bibitem[Ribeiro et al., 2022]{stacking1}
Ribeiro, M. H. D. M., da Silva, R. G., Moreno, S. R., Mariani, V. C., \& dos Santos Coelho, L. (2022). Efficient bootstrap stacking ensemble learning model applied to wind power generation forecasting. \textit{International Journal of Electrical Power \& Energy Systems}, 136, 107712.

\bibitem[Cao et al., 2023]{stacking2}
Cao, Y., Liu, G., Luo, D., Bavirisetti, D. P., \& Xiao, G. (2023). Multi-timescale photovoltaic power forecasting using an improved Stacking ensemble algorithm based LSTM-Informer model. \textit{Energy}, 283, 128669.

\bibitem[Nowotarski et al., 2016]{sister-forecast-load}
Nowotarski, J., Liu, B., Weron, R., \& Hong, T. (2016). Improving short term load forecast accuracy via combining sister forecasts. \textit{Energy}, 98, 40-49.

\bibitem[Liu et al., 2015]{sister-forecast}
Liu, B., Nowotarski, J., Hong, T., \& Weron, R. (2015). Probabilistic load forecasting via quantile regression averaging on sister forecasts. \textit{IEEE Transactions on Smart Grid}, 8(2), 730-737.

\bibitem[Moniz et al., 2018]{oversample1}
Moniz, N., Branco, P., \& Torgo, L. (2017). Resampling strategies for imbalanced time series forecasting. \textit{International Journal of Data Science and Analytics}, 3(3), 161-181.

\bibitem[Zhang et al., 2022]{oversample2}
Zhang, C., Fu, Y., \& Gong, L. (2022). Short-term electricity price forecast using frequency analysis and price spikes oversampling. \textit{IEEE Transactions on Power Systems}, 38(5), 4739-4751.

\bibitem[Vrbančič et al., 2020]{transfer_learning1}
Vrbančič, G., \& Podgorelec, V. (2020). Transfer learning with adaptive fine-tuning. \textit{IEEE Access}, 8, 196197-196211.

\bibitem[Kamath et al., 2019]{transfer_learning2}
Kamath, U., Liu, J., Whitaker, J., Kamath, U., Liu, J., \& Whitaker, J. (2019). Transfer learning: Domain adaptation. \textit{Deep learning for NLP and speech recognition}, 495-535.

\bibitem[Li \& Zhu, 2008]{LASSO}
Li, Y., \& Zhu, J. (2008). L 1-norm quantile regression. \textit{Journal of Computational and Graphical Statistics}, 17(1), 163-185.


\bibitem[Laha \& Rohatgi, 2020]{aggregate1}
Laha, R. G., \& Rohatgi, V. K. (2020). \textit{Probability theory}. Courier Dover Publications.

\bibitem[Zhang et al., 2016]{aggregate2}
Zhang, N., Kang, C., Singh, C., \& Xia, Q. (2016). Copula based dependent discrete convolution for power system uncertainty analysis. \textit{IEEE Transactions on Power Systems}, 31(6), 5204-5205.

\bibitem[Wang et al., 2018]{aggregate3}
Wang, Z., Wang, W., Liu, C., Wang, Z., \& Hou, Y. (2017). Probabilistic forecast for multiple wind farms based on regular vine copulas. \textit{IEEE Transactions on Power Systems}, 33(1), 578-589.

\bibitem[Ke et al., 2017]{LightGBM}
Ke, G., Meng, Q., Finley, T., Wang, T., Chen, W., Ma, W., ... \& Liu, T. Y. (2017). Lightgbm: A highly efficient gradient boosting decision tree. \textit{Advances in neural information processing systems}, 30.

\bibitem[Prokhorenkova et al., 2018]{CatBoost}
Prokhorenkova, L., Gusev, G., Vorobev, A., Dorogush, A. V., \& Gulin, A. (2018). CatBoost: unbiased boosting with categorical features. \textit{Advances in neural information processing systems}, 31.

\bibitem[Chen et al., 2021]{E2E_UC}
Chen, X., Yang, Y., Liu, Y., \& Wu, L. (2021). Feature-driven economic improvement for network-constrained unit commitment: A closed-loop predict-and-optimize framework. \textit{IEEE Transactions on Power Systems}, 37(4), 3104-3118.

\bibitem[Han et al., 2021]{E2E_ED}
Han, J., Yan, L., \& Li, Z. (2021). A task-based day-ahead load forecasting model for stochastic economic dispatch. \textit{IEEE Transactions on Power Systems}, 36(6), 5294-5304.

\bibitem[Sang et al., 2022]{E2E_ESS}
Sang, L., Xu, Y., Long, H., Hu, Q., \& Sun, H. (2022). Electricity price prediction for energy storage system arbitrage: A decision-focused approach. \textit{IEEE Transactions on Smart Grid}, 13(4), 2822-2832.


\bibitem[Browell et al., 2023]{data}
Browell, J., Haglund, S., Kälvegren, H., Simioni, E., Bessa, R., Wang, Y., \& van der Meer, D. (2023). Hybrid energy forecasting and trading competition. IEEE Dataport. https://dx.doi.org/10.21227/5hn0-8091

\bibitem[Breiman, 2001]{RF}
Breiman, L. (2001). Random forests. \textit{Machine learning}, 45, 5-32.

\bibitem[Makridakis et al., 2016]{M5}
Makridakis, S., Petropoulos, F., \& Spiliotis, E. (2022). Introduction to the M5 forecasting competition Special Issue. \textit{International journal of forecasting}, 38(4), 1279-1282.

\bibitem[REMIT, 2024]{REMIT}
Elexon, 2024. REMIT – Insights Solution. Available at: https://bmrs.elexon.co.uk/remit (accessed 10 April 2025).

\bibitem[Zhang et at., 2017]{regression-based test}
Zhang, H., Zhou, S., Zhang, K., \& Guan, J. (2017, February). Causal discovery using regression-based conditional independence tests. In \textit{Proceedings of the AAAI conference on artificial intelligence} (Vol. 31, No. 1).

\bibitem[Akiba et al., 2019]{Optuna}
Akiba, T., Sano, S., Yanase, T., Ohta, T., \& Koyama, M. (2019, July). Optuna: A next-generation hyperparameter optimization framework. In \textit{Proceedings of the 25th ACM SIGKDD international conference on knowledge discovery \& data mining} (pp. 2623-2631).


\end{thebibliography}
\end{document}